\documentclass{egpubl}
\usepackage{eg2025}
 
\ConferencePaper        %

\CGFStandardLicense

\usepackage[T1]{fontenc}
\usepackage{dfadobe}   
\usepackage{cite}  %
\usepackage{epigraph} 
\usepackage{amsmath} 
\usepackage{xspace}
\usepackage{pifont}
\usepackage{booktabs}
\usepackage{parskip}
\usepackage{pdfpages}
\BibtexOrBiblatex

\newcommand{\name}{FlairGPT\xspace}

\usepackage{tcolorbox}
\usepackage{xcolor}
\usepackage{soul} 
\usepackage{enumerate}
\usepackage{array}
\usepackage{multirow}

\definecolor{lightblue}{RGB}{173, 216, 230}
\definecolor{lightgreen}{RGB}{197, 224, 180}
\definecolor{lightorange}{RGB}{255, 230, 153}
\definecolor{lightpink}{RGB}{255, 204, 204}
\definecolor{lightyellow}{RGB}{255, 255, 204}
\definecolor{lightpurple}{RGB}{230, 190, 255}

\newtcolorbox{codefigure}[2][]{
    colback=white,
    colframe=darkgray,
    boxrule=0.5pt,
    arc=5pt,
    left=5pt,
    right=5pt,
    top=5pt,
    bottom=5pt,
    boxsep=0pt,
    title=#2,
    fonttitle=\large\bfseries,
    coltitle=white,
    colbacktitle=darkgray,
    before title={\vspace{5pt}},
    after title={\vspace{5pt}},
    #1
}

\newcommand{\hlc}[2]{\sethlcolor{#1}\hl{#2}}

\electronicVersion
\PrintedOrElectronic

\graphicspath{ {paper/figures/png_figures/} }

\title[\name: Repurposing LLMs for Interior Designs]%
      {\name: Repurposing LLMs for Interior Designs}

\author{
    \Large{\href{https://flairgpt.github.io/}{flairgpt.github.io}}\\
    \large{
    Gabrielle Littlefair\textsuperscript{1} \hspace{5mm}
    Niladri Shekhar Dutt\textsuperscript{1} \hspace{5mm}
    Niloy J. Mitra\textsuperscript{1,2}} 
    \\
    \large{\textsuperscript{1}University College London \hspace{5mm}
    \textsuperscript{2}Adobe Research
    }
}

\begin{document}

\teaser{
 \centering
 \includegraphics[width=\linewidth]{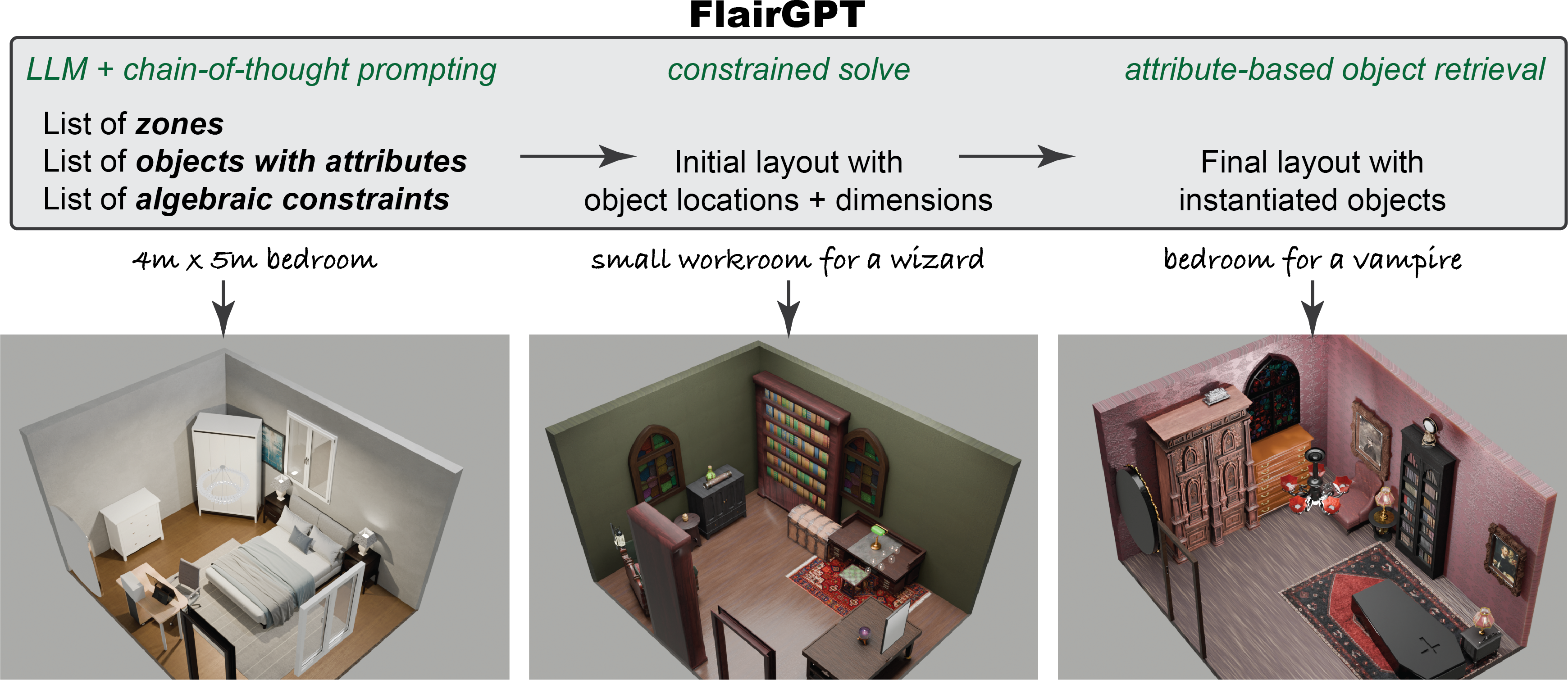}
  \caption{
  We investigate if large language models (LLMs) can be used as interior designers. We show that LLMs can be systematically probed and combined with traditional optimization to produce aesthetically-pleasing and functional interior designs. In these examples, our method \name, starting from text probes, produces the final layouts including object selection, their placement, as well as their styles. }
\label{fig:teaser}
}

\maketitle
\author{Gabrielle Littlefair, Niladri Shekhar Dutt, Niloy J. Mitra}

\begin{abstract}
Interior design involves the careful selection and arrangement of objects to create an aesthetically pleasing, functional, and harmonized space that aligns with the client's design brief. This task is particularly challenging, as a successful design must not only incorporate all the necessary objects in a cohesive style, but also ensure they are arranged in a way that maximizes accessibility, while adhering to a variety of affordability and usage considerations.
Data-driven solutions have been proposed, but these are typically room- or domain-specific and lack explainability in their design design considerations used in producing the final layout. 
In this paper, we investigate if large language models~(LLMs) can be directly utilized for interior design.
While we find that LLMs are not yet capable of generating complete layouts, they can be effectively leveraged in a structured manner, inspired by the workflow of interior designers. By systematically probing LLMs, we can reliably generate a list of objects along with relevant constraints that guide their placement.
We translate this information into a design layout graph, which is then solved using an off-the-shelf constrained optimization setup to generate the final layouts.
We benchmark our algorithm in various design configurations against existing LLM-based methods and human designs, and evaluate the results using a variety of quantitative and qualitative metrics along with user studies. 
In summary, we demonstrate that LLMs, when used in a structured manner, can effectively generate diverse high-quality layouts, making them a viable solution for creating large-scale virtual scenes. 
Code is available via the \href{https://flairgpt.github.io/}{project webpage}.

\ccsdesc[500]{Computing methodologies~Shape analysis}
\ccsdesc[300]{Computing methodologies~Natural language processing}
\ccsdesc[100]{Computing methodologies~Machine learning}
\printccsdesc

\end{abstract}  

\section{Introduction}
\label{sec:intro}
\epigraph{We shape our homes and then our homes shape us.}{Winston Churchill}
Interior designing is the art of creating balanced, functional, and aesthetically pleasing spaces based on intended space usage and adjusted to individual preferences. The goal is to propose a selection of objects, both in terms of the type and style of the objects along with their arrangement, that best serves the project brief provided by the client. A good design not only considers the aesthetic look of the objects, but also factors in the flow of the designed space, taking into consideration affordability of the objects along with their functionality and access space.

The design task is challenging, as one has to balance aesthetics, functionality, and practicality within a given space while considering the user's needs, preferences, and budget. It is particularly difficult to identify, keep track, and balance a variety of conflicting constraints that arise from ergonomics and usage while harmonizing furniture, lighting, and materials. Hence, users often take shortcuts and fall back to a rule-based or preauthored solution that best fits their specifications. However, achieving a customized, cohesive, visually appealing and functional design requires creativity, technical expertise, and remains difficult for most users.

To gain inspiration, we first studied how interior designers approach the problem. Upon receiving project briefs, they divide the space into \textit{zones} according to their intended function. They then begin by selecting and placing the \textit{focal} objects for the key zones, before arranging other objects around them. Throughout this process, they carefully consider design aspects to ensure that objects are easily accessible and usable and that the room has \textit{good flow} to facilitate movement. Finally, they incorporate lighting and decide on the style of the objects, as well as the wall and floor style to create a harmoniously designed space. The most non-trivial aspect is the variety of spatial and functional considerations that designers consider and conform to while designing the space.

\begin{figure}[b!]
    \centering
\includegraphics[width=\columnwidth]{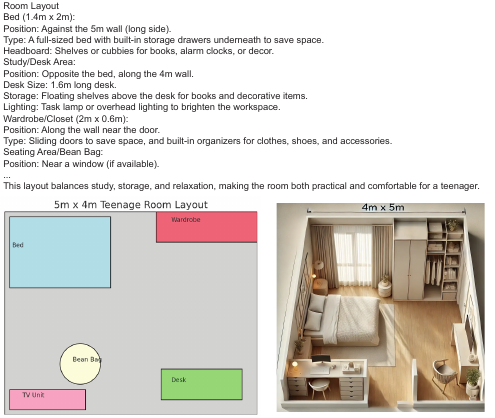}
    \caption{Layouts Generated by ChatGPT~\cite{chatgpt}. (Top)~Directly querying LLMs to generate room layouts yields useful guidance but not a floorplan. (Bottom-left)~Asking for a floorplan results in an overly simplistic one, with very few objects and impractical proportions—such as a TV unit nearly as long as the bed. Additionally, essential objects, like a chair for the desk, are missed. (Bottom-right)~When prompted to generate design images, the results, while aesthetically pleasing, are often functionally impractical, as shown in the image on the right. For instance, the desk and chair are incorrectly oriented, rendering the chair inaccessible.}
    \label{fig:motivation}
\end{figure}

In this paper, we ask if large language models~(LLMs) can be repurposed for interior design. 
We hypothesize that LLMs that have been trained on various text corpora, including design books and blogs, are likely to know about layout design concepts. We ask how explicit these concepts are and how good they are in quality. 
Directly querying LLMs to produce room layouts based on text guidance (e.g., `Please design a drawing room of size 4m$\times$5m for a teenager who loves music') regularly produced mixed results that had good design ideas but not usable in practice (see \autoref{fig:motivation}). Although the output images of the room looked aesthetically pleasing, closer inspection revealed many design flaws. Unfortunately, when asked for output floorplans, LLMs produced rather basic layouts that did not meet expectations. 

Interestingly, we found that LLMs have good knowledge of individual design considerations, including non-local constraints. For example, when asked about `the most important design consideration for a kitchen' LLMs described the \textit{kitchen work triangle}, which is an important design consideration that many of us are unaware of and can easily get wrong, severely affecting the functionality of the space. Encouraged by this and inspired by interior designers' workflow, we break the interior design task into stages. Instead of directly using LLMs to get the final layout, we progressively probe the LLMs, in a structured fashion, to first zone the given space and then extract a list of objects to populate the different zones. More importantly, we also elicit a list of intra-object and inter-object constraints along with descriptive attributes for the selected objects. Then, using a symbolic translation, we organize the LLMs output into a layout constraint graph by converting the textual constraints to algebraic constraints in terms of the object variables (i.e., their size and placement). We then obtain the layout by solving the resultant constrained system. Finally, we retrieve objects to populate the designed layout using the object-specific types and attributes obtained from the LLMs to produce the final layouts. 

This paper introduces \name, a novel framework for generating functional and aesthetic room layouts by leveraging LLMs in combination with optimization techniques. The contributions of our approach are as follows:
\begin{enumerate}
\item \name proposes an approach, grounded in interior designers’ workflows, that links structured LLM queries, a library of objective functions augmented with language doc-strings, and traditional optimization via inferred constraint graphs to produce room layouts.
\item The generated layouts are functional, editable, and interpretable.
\end{enumerate}

\autoref{fig:teaser} presents a selection of example outputs from our method, \name. 
We evaluated our method in a variety of interior design settings. We compared ours with the latest interior design alternatives (e.g.,  ATISS~\cite{ATISS}, Holodeck~\cite{holodeck}, LayoutGPT~\cite{layoutgpt}, I-Design~\cite{idesign}) and against user-designed layouts. We compared the quality of our designs and those produced by competing methods using  different user studies. Users consistently preferred our generations over the others, including those done by novice users, and scored ours well with respect to adhering to design specifications as well as producing functionally useful layouts.
We also evaluate perform quantitative evaluation on the generated layouts. 
In addition, we report our findings on the aspects of the design process where LLMs offer significant value and those that are best managed, at least for now, by human expertise. %

\section{Related Works}
\label{sec:related}
\paragraph*{Optimization-based layouts.}
Interior design relies on spatial arrangement, human-centric aesthetics, and functional optimization~\cite{alexander2018pattern}. Early computational approaches for generating simple layouts~\cite{harada,michalek2002interactive} concentrated on manually defining local constraints and employing optimization techniques to solve for optimal spatial arrangements. Later, inspired by established interior design guidelines, Merell et al.~\cite{merell} introduced an interactive system that allowed users to define the shape of the room and a selected set of furniture, after which the system generates design layouts that adhere to specified design principles. Make it home~\cite{makeithome} employed hierarchical and spatial relationships for furniture
objects with ergonomic priors in their cost function to yield more realistic furniture arrangements. 
In a recent optimization method, Weiss et al.~\cite{fast_scalable} use physics-based principles to create room layouts by treating objects as particles within a physical system. The method emphasizes both functionality and harmony in the room by applying specific constraints to ensure walkways, maintain balanced visual appeal around a focal point, etc. However, it still requires users to manually specify constraints. 

\paragraph*{Data-driven layouts.}
Rather than relying on hard coded rules for optimization, modern data-driven methods aim to learn such concepts automatically~\cite{ritchie2019fast,wang2018deep,tang2023diffuscene}. For example, ATISS~\cite{ATISS} treats indoor scene synthesis as an unordered set generation problem, to allow flexibility by avoiding the constraints of fixed object orderings. ATISS uses a transformer architecture to encode floorplans and object attributes to sequentially place objects based on category, size, orientation, and location. While visually appealing, ATISS suffers from practical limitations such as overlapping objects. To enhance practicality, LayoutEnhancer~\cite{layoutenhancer} integrates expert ergonomic knowledge—such as reachability, visibility, and lighting—directly into the transformer model for indoor layout generation. However, the method falls short in considering stylistic elements, limiting its ability to generate complex aesthetically tailored designs. SceneHGN~\cite{scenehgn} creates a hierarchical graph of the scene to capture relationships among objects to produce visually coherent 3D environments. Tell2Design~\cite{tell2} reformulates the task of generating floor plans as a sequential task where the input is language instructions and the output is bounding boxes of rooms. Although data-driven methods can produce good results, they are limited in diversity and creativity due to their reliance on curated datasets and are often restricted to special types of rooms and/or objects. 

\paragraph*{LLM-based layouts.}
With advances in capabilities of Large Language Models~\cite{gpt3,gemini,touvron2023llama, jiang2024mixtral}, LLMs are being increasingly used to solve a plethora of complex tasks such as reasoning~\cite{reasoning}, programming~\cite{codellama}, discovering mathematical concepts~\cite{romera2024mathematical}, conducting scientific research~\cite{lu2024ai}, etc. Building on this success, the integration of LLMs in scene synthesis offers the ability to generate context-aware designs by interpreting and applying textual descriptions directly to the synthesis process. This enables a more dynamic and flexible approach, allowing for the integration of complex design principles that are often difficult to encode through conventional algorithms. 

Holodeck~\cite{holodeck} utilizes LLM to expand user text prompts to generate a scene into actionable scene elements. However, the actual placement and relationship of objects are governed by a set of predefined spatial rules hard-coded into the system that can limit the flexibility and creativity of the system to adapt to unconventional or complex designs. In a very recent system, LayoutGPT~\cite{layoutgpt} uses LLMs to generate scene layouts by treating elements within the scene as components that can be described and adjusted programmatically akin to web elements in CSS. %
In another notable effort, Aguina-Kang et al.~\cite{open_universe} employ LLMs to create more detailed scene specifications from simple prompts, identify necessary objects and finally generate programs in domain specific language to place those objects in the scene. After establishing one of ten relationships between objects from a library, the final placement is obtained using gradient descent based optimization. LLplace~\cite{llplace} fine tunes Llama3~\cite{touvron2023llama} on an expanded 3D-Front Dataset~\cite{3dfront} to allow users a more interactive way to add and remove objects in a conversational manner. I-Design~\cite{idesign} uses multiple LLMs to convert a text input into a scene graph and obtain a physical layout using a backtracking algorithm. Strader et al.~\cite{ontologies} leverage LLMs to build ``spatial ontology'' (to store concepts), which is used in node classification systems of 3D scene graphs. 

In contrast to other LLM and optimization methods, such as I-Design, which employs topological sorting to establish hierarchical ordering on the graph, our approach aligns more closely with artist workflows and zoning principles. This enables a natural and interpretable ordering process, which allows faithfully capturing the complexity of spatial relationships and functional constraints.
More importantly, our method effectively combines LLM with cost-based optimization, further augmented by the use of language doc-strings. This integration not only provides a controllable framework for ensuring functionality but also achieves superior results in terms of feasibility. Our evaluation (Section~\ref{sec:evaluation}) demonstrates the effectiveness of this approach, surpassing both Holodeck's constraint-based layout design~\cite{holodeck} and against I-Design's backtracking strategy~\cite{idesign}.

\begin{figure*}
    \centering
\includegraphics[width=\linewidth]{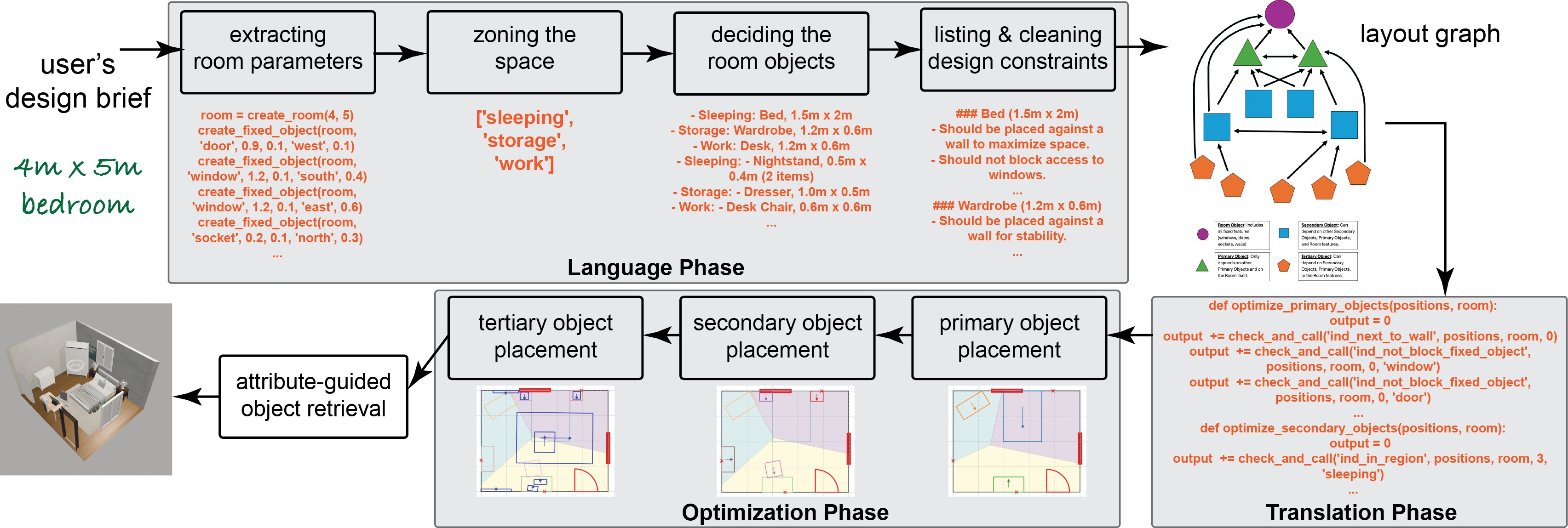}
    \caption{Method overview. \name begins by taking the user's design request as a text prompt and querying an LLM to extract key room parameters, such as dimensions and the location and number of windows, doors, and sockets. Next, following a designer’s workflow, the LLM generates an ordered list of zones, specifying the functional purpose of different areas within the room. Based on these zones, a prioritized list of required objects is generated, complete with descriptions and dimensions. These objects serve as the nodes of a layout graph, with inter- and intra-object constraints—defined by the LLM—forming the edges. The natural language constraints provided by the LLM are translated into algebraic forms by querying the LLM to map these constraints to a predefined library of cost functions. Once these cost functions are established, the placement and orientation of objects are progressively optimized according to their hierarchical importance. Finally, objects are retrieved, based on their descriptions, and incorporated into the scene.}
    \label{fig:overview}
\end{figure*}

\section{Design Considerations}
\label{sec:design_overview}

In this section, we briefly summarize the process followed by interior designers as documented in design literature books~\cite{Brooker2013,Mitton2012,alexander2018pattern}. 

The process starts with a \textit{design brief} where the clients describe how they plan to use the space, provide background on their preferences, and detail the current layout of the space (e.g., walls, doors, windows). Budget and time frames are also discussed in this stage, but we ignore these in our setup.

\textit{Space planning}, the next phase, is the most challenging. This involves creating functional layouts and optimizing the use of space. Specifically, they determine the choice and arrangement of furniture while considering flow, accessibility, and ergonomics. Designers typically start by collecting measurements of the space and noting the features of the room such as doors, windows, and electrical outlets. 
Next, they \textit{zone} the space by partitioning the region into distinct areas based on its functions. For example, in an open-plan layout, designers allocate areas for dining, working, and socializing without the need for physical barriers. In this stage, they also take traffic flow into account to create pathways or circulation areas that avoid overcrowding and allow a smooth transition between zones. 
Having zoned the space, designers then select and place key pieces of furniture, usually referred to as \textit{primary objects}, in strategic positions. Large items (e.g., sofas, tables, beds) are first positioned in order to anchor the space. Designers use their experience to  balance functionality and aesthetics to create visual interest and harmony in the space.
Next, they incorporate \textit{secondary objects} (such as chairs, appliances, etc.) around the primary objects to ensure the regions are functional. At this point, artificial lighting is also added if necessary. Besides selecting the types and sizes of objects, designers also consider their color and finish to create a cohesive look in the designed space while maintaining its functionality.

Finally, during \textit{design development}, designers collect client feedback based on previsualization of the space and iterate on the design to better align the space to their clients' vision. 

\section{Algorithm}
\label{sec:algorithm}

Our method consists of three key phases. In the first phase, the \textit{Language Phase}, we progressively query the LLM to make informed decisions about the room’s layout and design. The model identifies all relevant objects for the space along with their dimensions (width and length). More importantly, the LLM provides a set of spatial constraints that governs the positioning and arrangement of these objects. In the second phase, the \textit{Translation Phase}, we convert the language-based constraints obtained from the LLM into executable function calls, drawing from a predefined library of constraint cost functions thus forming a layout constraint graph. 
Finally, in the \textit{Optimization Phase}, we use an optimization~(SLSQP) to find a minimal-cost solution that satisfies the combined set of constraints. We stagger this phase into multiple iterations with random initial configurations. Upon completion, we obtain the full specification of all objects, including their style, dimensions, positions, and orientation.
We now provide details on each phase. The exact LLM query formulation is available in the supplemental material.

\subsection{The Language Phase}
\paragraph*{User input.}
We expect the user to provide a textual description of the room they wish to generate. This input can range from simple prompt, such as “a bedroom,” to more detailed specifications like, “a $5\times5$m bedroom for a young girl who enjoys painting while looking out of her window.” This flexibility allows users to define a wide variety of room configurations.

\paragraph*{A. Extracting room parameters.}
Once the user input has been provided, we query the LLM to establish the fundamental parameters of the room that serve as the fixed boundary condition for the rest of the stages. The model generates the dimensions of the room (width and length), with the height fixed at 3 meters by default. The LLM also prescribes how many windows, doors, and electrical sockets the room requires, as well as their placements (which wall they should be on and their horizontal position along that wall). Additionally, the model provides the width of the windows and doors. 
Note that we designed a fixed schema to convert user specifications to queries for the LLM. Please see supplemental for details. 
Users can alternatively bypass this step if they prefer to directly input the room specifications.

\paragraph*{B. Zoning the space.}
Next, similar to how human designers proceed, we query the LLM to determine the core purposes of the room, which define its \textit{zones}. The number and type of zones vary depending on the room’s size and intended use. The LLM outputs an ordered list of zones, ranked by significance. For example, in a bedroom, the zones can include \{\texttt{sleeping}, \texttt{storage}, \texttt{dressing}\} areas. We denote this ordered list by $\mathcal{Z}:=\{z_1,\dots,z_k\}$. Note that we do not partition the room into zones at this stage. 

\paragraph*{C. Deciding the room objects.} Our next major design task is to decide which objects to include in the room along with their size and textual description. Again, following designers' workflow, we proceed in stages. 

\noindent{\em (i)~Listing the  \textbf{primary objects}.} We define a primary object as the most essential object required for each zone to fulfill its intended purpose (these are often referred as \textit{focal} objects). Again, we query the LLM to determine the primary objects, along with their dimensions (see supplemental for query schema). The output is an ordered list where each entry includes the primary object $p_i$ corresponding to zone $z_i$, as well as the object's width ($w_i$) and length ($l_i$). Thus, the list of primary objects takes the form 
$\mathcal{P}:=\{
(p_1, w_1, l_1),\dots,
(p_k, w_k, l_k)
\}.
$
So far, we have obtained the type, width, and length for each primary object, but not their height.

\begin{figure}[b!]
    \centering
    \includegraphics[width=\linewidth]{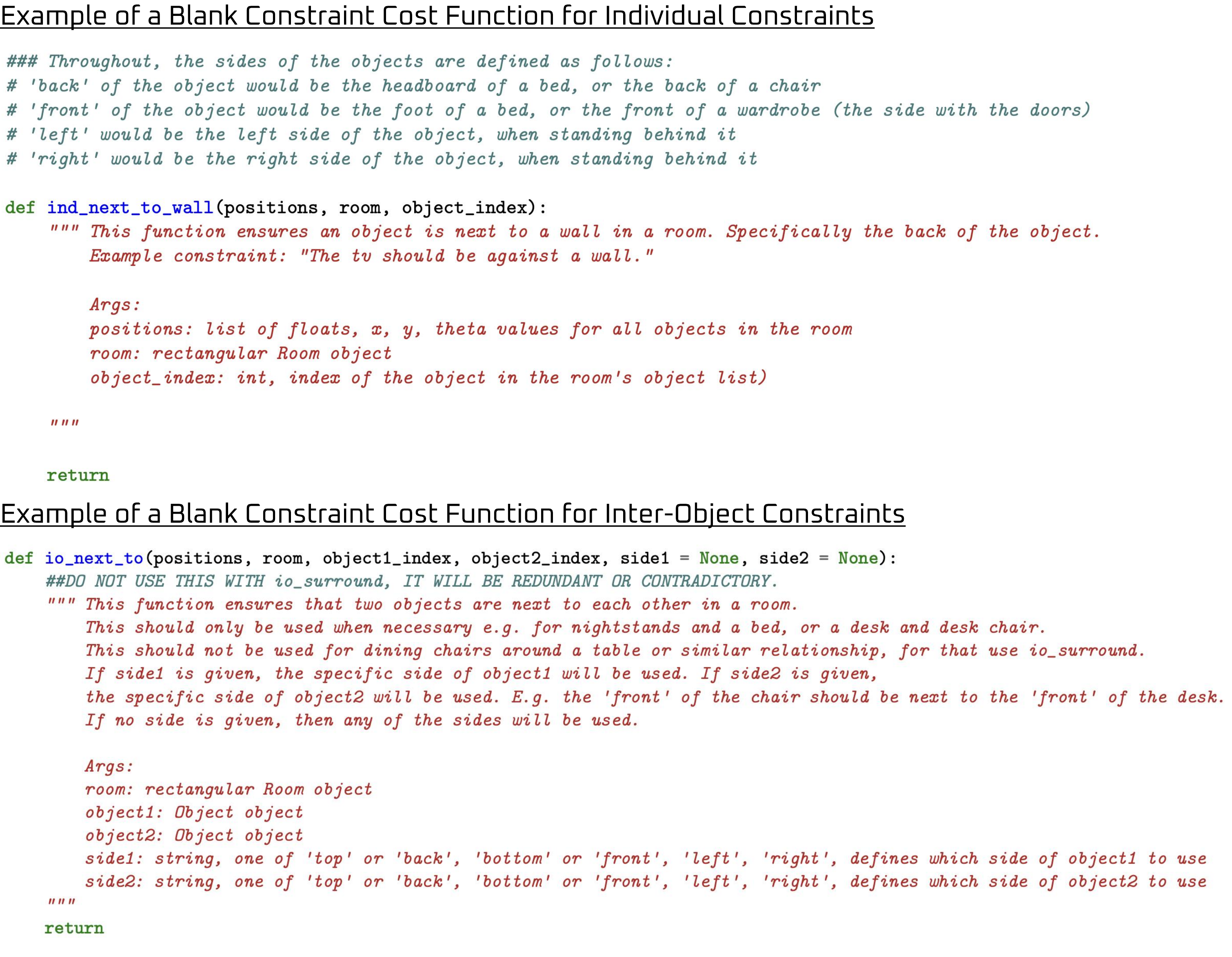}
    \caption{Doc string for library functions. An example of our \texttt{docstrings} which contain usage examples and thorough descriptions of each function’s purpose and its parameters. Note that the underlying implementation of the functions is absent. The LLM is tasked to map each language-based constraint to a corresponding cost function within the library during the language phase.}
    \label{fig:docstring}
\end{figure}

\begin{figure*}
    \centering
\includegraphics[width=\linewidth]{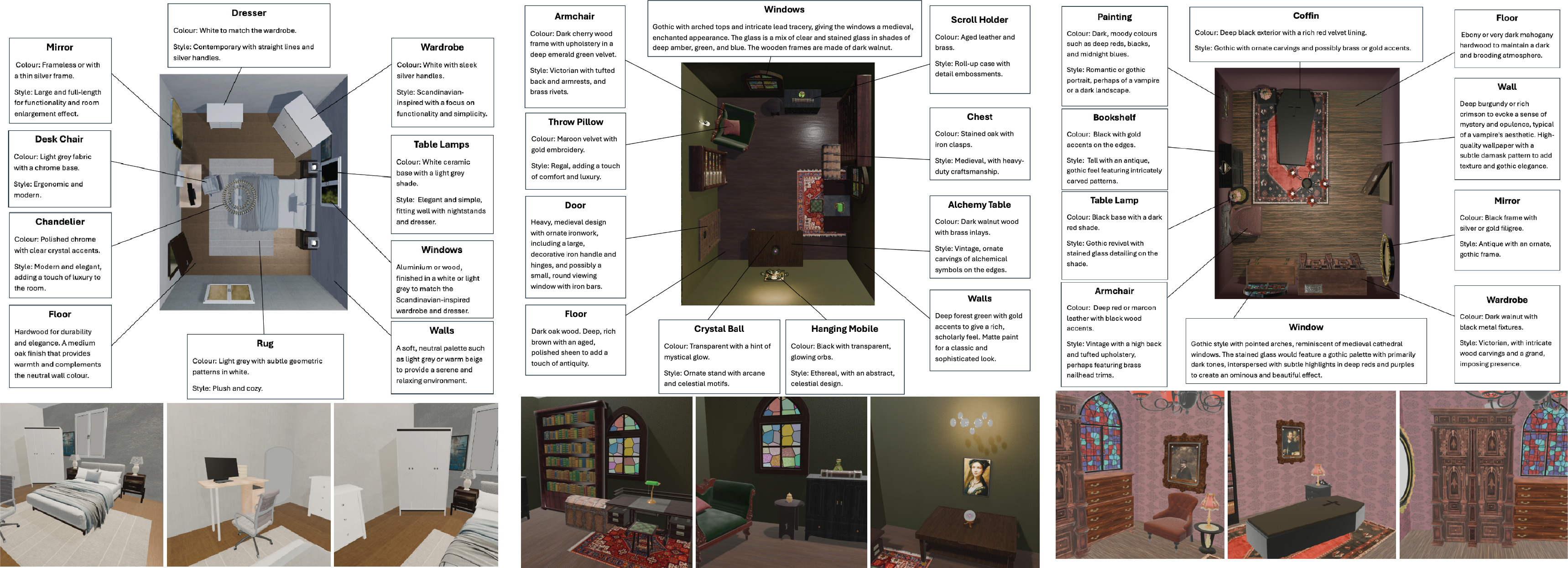}
    \caption{Generated layouts by \name. We present varied layouts designed by \name for three distinct prompts (from left to right)- ``4m x 5m bedroom'', ``small workroom for a wizard'', and ``bedroom for a vampire''. Alongside each layout, we include descriptions of selected objects provided by the LLM, which closely align with the user's design brief. Notably, \name makes creative and context-appropriate object choices, such as a scroll holder and a crystal ball for a wizard's workroom, and a coffin in place of a traditional bed in the case of a vampire's bedroom, reflecting the thematic style of the input prompts.}
    \label{fig:teaser_legend}
\end{figure*}

\noindent{\em (ii)~Listing the  \textbf{secondary objects}.}
We then query the LLM to identify secondary objects, defined as additional items that enhance the functionality of each zone, provided they are floor-based (excluding rugs). The output is an ordered list of secondary objects $\mathcal{S}$, where each object $s_i$ includes its width ($w_i$), length ($l_i$) and the corresponding zone $z(s_i)$ to which it belongs. Note that each zone can have multiple secondary objects. In addition, the output specifies how many of each object are needed. 
For example, four dining chairs or two nightstands might be required in a given zone.
Thus, we have, 
$\mathcal{S}:=\{
(s_1, w_1, l_1, z(s_1)),\dots,
(s_{n_s}, w_{n_s}, l_{n_s}, z(s_{n_s}))
\}$ with ${n_s}$ being number of secondary objects. 

\noindent{\em (iii)~Listing the  \textbf{tertiary objects}.}
We then query the LLM to generate the final set of objects, the tertiary ones. Such objects are `attached' to specific primary/secondary objects or room boundary. These include ceiling-mounted objects (e.g., chandeliers), wall-mounted objects (e.g., paintings), objects placed on surfaces (e.g., table lamps), and rugs. While majority of the tertiary objects are decorative, functional items such as computers and lighting can also be suggested at this stage. We also query the LLM for detailed placement instructions, specifying how and where these objects should be positioned relative to other objects or zones within the room. For example, the LLM might suggest, “place a painting on the wall above the bed.”

The output is an unordered list $\mathcal{T}$ of tertiary objects, each described in relation to another object (either Primary or Secondary), a boundary wall, or simply a specific zone. For each tertiary object $t_i$, we also obtain its type ($\text{type}_i$), one of wall, floor, ceiling, or surface, along with its width ($w_i$), length ($l_i$), and a language-based placement constraint ($c_i$). The final output is an unordered list
$\mathcal{T}:=\{(t_1, w_1, l_1, \text{type}_1, c_1), \dots, 
(t_{n_t}, w_{n_t}, l_{n_t}, \text{type}_{n_t}, c_{n_t})
\}$, where $\text{type}_i$ specifies the object type, $c_i$ provides the placement instructions, and ${n_t}$ being the number of tertiary objects. 

The language constraints ($c_i$) are treated separately from those of primary and secondary objects, as tertiary objects can be positioned in ways that others cannot — such as on the ceiling, walls, atop other objects, or underneath primary or secondary objects.

\noindent{\em (iv) Determining style for the objects.}
Having listed all the objects, we move on to determine the style of the room and the individual objects using the given description for the room. We  query the LLM to specify the style of the room and each individual object. The LLM provides textual details such as materials, colors, and patterns for the walls and floors. For instance, it might suggest “dove grey paint with an accent wall featuring a subtle geometric wallpaper.” Each object, including windows and doors, is further described by the LLM in terms of material, color, and overall aesthetic.

\subsubsection*{D. Listing of design constraints}

So far we have the specification of the room boundary and a textual list of the objects to be placed in the room. The list of objects $\mathcal{P} \cup \mathcal{S} \cup \mathcal{T}$ forms the nodes of our layout graph. Next, we use the LLM to list all the relevant inter- and intra-object constraints, which become the (undirected) edges of our layout graph. We only consider pairwise constraints in our setup. 

\noindent{\em (i) Intra-object constraints.} These constraints refer to those that apply to a single object, either a primary or secondary object, and any features of the room (including walls, windows, doors, and sockets). These constraints govern the positioning and usability of an individual object. For example, the LLM might specify, “the bed should have its headboard against the wall,” or “the bed should not be too close to a window to avoid drafts.” This category also includes accessibility requirements, such as determining which sides of the object must remain accessible for it to function properly. At this stage, we query the LLM to generate all such constraints by looping over all the nodes in $\mathcal{P} \cup \mathcal{S}$ and collect them for all the primary and secondary objects in natural language.

\noindent{\em (ii) Inter-object constraints.}
These constraints involve relationships between pairs of primary and secondary objects. For instance, the LLM might suggest, “the mirror should not face the bed,” or “the bed should be placed between the two nightstands.” When the constraint applies only between primary objects, we encourage the LLM to create simple spatial relationships such as “near to” or “far from,” since these objects often belong to different zones. 

\noindent{\em (iii) Constraint cleaning.}
The final step in the Language Phase serves as a self-correction tool. We query the LLM to review and refine the generated constraints. This involves merging any similar constraints, removing duplicates, and simplifying the constraints into more straightforward language to minimize errors during the Translation Phase. The LLM also identifies and eliminates any contradictory constraints. Additionally, we use the LLM to split constraints that contain multiple pieces of information. For example, “the bed should not block windows or doors” would be split into “the bed should not block windows” and “the bed should not block doors”. This is not applied to the tertiary constraints, due to there only being one constraint per tertiary object.

{\scriptsize
\begin{codefigure}{Example of language constraints being translated into code}
\begin{minipage}{\linewidth}

\#\#\# 1. Bed (1.6m x 2m)

\hlc{lightblue}{- Should be placed against a wall for headboard support.}

\hlc{lightgreen}{- Should not block any doors or windows.}

\hlc{lightorange}{- Should have clear access on at least two sides for ease of getting in and out (usually the longer sides).}

\hlc{lightpink}{- Ensure there are no electrical sockets directly behind the head area to avoid tangling cords.}

- Ideally, position it such that there is space for nightstands on either side.

\vspace{0.5em}

{\ttfamily
def optimize\_primary\_objects(positions, room):\\
\hspace*{2em}output = 0\\
\hspace*{2em}\hlc{lightblue}{output += ind\_next\_to\_wall(positions, room, 0)}\\
\hspace*{2em}\hlc{lightgreen}{output += ind\_away\_from\_fixed\_object(positions, room, 0, fixed\_object\_type='door')}\\ 
\hspace*{2em}\hlc{lightgreen}{output += ind\_away\_from\_fixed\_object(positions, room, 0, fixed\_object\_type='window')}\\
\hspace*{2em}\hlc{lightorange}{output += ind\_accessible(positions, room, 0, sides=['bottom', 'left'])}\\
\hspace*{2em}\hlc{lightpink}{output += ind\_away\_from\_fixed\_object(positions, room, 0, fixed\_object\_type='socket', min\_dist=2.0)}\\
\hspace*{2em}\hlc{lightyellow}{output += in\_bounds(positions, room)}\\
\hspace*{2em}\hlc{lightyellow}{output += no\_overlap(positions, room)}\\
\hspace*{2em}\hlc{lightyellow}{output += aligned(positions, room)}\\
\hspace*{2em}\hlc{lightyellow}{output += balanced(positions, room)}\\
\hspace*{2em}\hlc{lightyellow}{output += wall\_attraction(positions, room)}\\
\hspace*{2em}return output
}

\end{minipage}
\end{codefigure}
}

\subsection{The Translation Phase}

Next, we convert the language constraints into algebraic forms. 
For this phase, we created a “blank” version of our library of constraint cost functions. This blank version contains the names of the functions, along with detailed \texttt{docstrings} for each function. These \texttt{docstrings} include usage examples and thorough descriptions of each function’s purpose and its parameters. Note that these strings only provide function names and lists of variables to the LLMs, but not the underlying implementation of the functions. \autoref{fig:docstring} shows an example; more details are provided in the supplemental. 

The purpose of these blank functions is to utilize the natural language processing capabilities of the LLM to map each language-based constraint to a corresponding cost function within the library. This process is carried out in three distinct stages: one for Individual or Intra-Object constraints, one for for Inter-Object constraints, and one for tertiary constraints. By processing these constraints separately, we ensure the correct type of function is applied, reducing the risk of using the wrong function for a particular constraint.

If no suitable matching function can be found for a given constraint, we discard the corresponding language constraint. Additionally, if the parameters provided to the function do not match the expected inputs, we ensure the function safely returns a cost value of 0, reducing errors in the subsequent optimization process.

\subsection{The Optimization Phase}

Finally, we are ready to place the objects by determining the coordinates $(x, y)$ of the centroid and the orientation $(\theta)$ of each object. Given the highly constrained nature of the problem, we split the optimization process into several steps, progressively solving for the layout. For each step, we compute a combined cost using all relevant constraint cost functions, as provided by our library functions, and find the optimal solution using a Sequential Least Squares Quadratic Programming (SLSQP) solver. To improve robustness, we repeat each optimization with different initializations for the variables, and take the best solution. For each object $i$, we optimize its position $(x_i, y_i)$ and orientation $\theta_i$. Note that we define orientation with respect to forward-facing direction of each object. 

We include five additional cost functions for the first two stages of the optimization (i.e., primary and secondary object placement) as they fit most layouts well. We note that these fixed cost functions are substantially augmented by the LLM during the translation phase by choosing applicable cost functions from a library (via doc-strings) of different loss functions along with added suitable constraints depending on the layout. We describe our fixed cost functions below:

\begin{enumerate}[(i)]
\item A no-overlap cost $C_{\text{over}}$ which penalizes intersections between objects. In \autoref{eqn: no_overlap}, we show the formulation where, for every pair of objects $i$ and $j$, we find the projected 2D polygon formed by their intersection ($\text{poly}_{ij}$). We then apply a function $f$, which sums the squared lengths of the sides of this polygon. This calculation is also applied to every object $i$ in relation to any doors $d$, ensuring that no object intersects with a door, and for this term we add a scaling factor $\lambda_1$ (we use 100 in our experiments). In particular, the cost term is as follows, 
\begin{equation}\label{eqn: no_overlap}
C_{\text{over}} := \sum_i [\sum_{j > i} f(\text{poly}_{ij}) + \lambda_1 \sum_d f(\text{poly}_{id})].
\end{equation}

\item An in-bounds cost $C_{\text{bound}}$ which penalizes objects that extend beyond the room’s boundaries. In \autoref{eqn:in_bounds}, we show the formulation for object $i$, where we iterates over its corners $c_{ij}$; $I_{c_{ij}}$ is an indicator variable that takes a value of 1 if the corner $c_{ij}$ lies within the room boundary, $B$, otherwise it is 0. 
\begin{equation}\label{eqn:in_bounds}
C_{\text{bound}, i} := \sum_{j = 0}^3 (1 - I_{c_{ij}})  \text{ dist}(c_{ij}, B)^2.
\end{equation}

\item An alignment cost $C_{\text{align}}$ which weakly penalizes orientations that deviate from the cardinal directions. Namely, we use
\begin{equation}\label{eqn:aligned}
    C_{\text{align}, i} := \frac{\sin^2(2\theta_i)}{5}. 
\end{equation}

\item A balanced placement cost $C_{\text{bal}}$ that penalizes deviations of the weighted centroid of all of the objects from the center of the room. The formulation of this is shown in \autoref{eqn:balanced}, where $w$ and $l$ are the width and length of the room, and for object $i$, $a_i$ is the area of the bounding box. 

\begin{equation}\label{eqn:balanced}
   C_{\text{bal}} := \left( \frac{\sum_i a_i x_i}{\sum_i a_i} - \frac{w}{2}\right)^2 + \left(\frac{\sum_i a_i y_i}{\sum_i a_i} - \frac{l}{2} \right)^2.
\end{equation}

\item A wall-attraction cost $C_{\text{wall}}$ which weakly encourages objects to be near the walls. This is to prevent `floating' objects from being placed centrally in the room. The formulation is shown in \autoref{eqn:wall_attraction}, where if the distance of object $i$, $o_i$, from the closest wall is greater than a given threshold $T$, a penalty is applied. We find that scaling this cost with a factor ($\lambda_2$) works better. We use $\lambda_2 = 20$ in all of our experiments.
\begin{equation}\label{eqn:wall_attraction}
   C_{\text{wall}, i} := \frac{1}{\lambda_2} \left( \min(T - \min_{\omega \in {\text{walls}}} \text{dist}(o_i, \omega), 0.0)^2\right).
\end{equation}

\end{enumerate}

These functions account for all objects that are present in the room at the time of optimization. For instance, during the first optimization step, only overlaps between the primary objects are considered. Subsequently, intersections involving the newly added secondary objects are evaluated, along with any intersections between the secondary and previously placed primary objects.

\paragraph*{A. Primary object placement.}
We begin by optimizing the locations and orientations of the primary objects ($i \in \mathcal{P}$). These locations and orientations are influenced by room features such as walls, windows, doors, and sockets, as well as by the positioning of other primary objects. We solve the following SLSQP, where $\lambda_i, i \in {1, 2, 3}$ are tunable parameters. We use $\lambda_3 = 5, \lambda_4 = 10$ and $\lambda_5 = 10$ in all of our experiments.

\begin{equation}\label{eqn:pri_place}
\begin{aligned}
\min_{\{x_i, y_i, \theta_i\}_{\mathcal{P}}} C_{\text{pri}} := \lambda_3 C_{\text{over}} + \lambda_4 C_{\text{bal}} + \sum_{i \in \mathcal{P}} ( C_{\text{lang}, i} + \\ \lambda_5 C_{\text{bound}, i} +  C_{\text{align}, i} +   C_{\text{wall}, i} )
\end{aligned}
\end{equation}

Once the positions and orientations are determined, we initialize the zones, setting each initial centroid as the position of the corresponding primary object's. We then use Voronoi segmentation based on these centroids to define the corresponding zones ($z_i)$.

After optimizing the primary objects, we record the name, width, length, style description, coordinates of its centroid, and orientation  $(p_i, w_i, l_i, \text{style}_i, x_i, y_i, \theta_i)$ of each object. These values are held fixed during subsequent optimizations.

\paragraph*{B. Secondary object placement.}
At this stage, the initial zones have been defined, and the positions and orientations of the primary objects are fixed. We then proceed zone by zone, to add the secondary objects ($i \in \mathcal{S}$). The positioning and orientation of these secondary objects are influenced by room features (such as walls, windows, doors, and sockets), the primary objects, and other secondary objects. We carry forward any accessibility constraints from the first stage, in order to ensure that the primary objects remain accessible. We add a default constraint $C_{\text{zone}}$ with a scaling factor $\lambda_6$ (we use $\lambda_6 = 10$ in all our experiments) to ensure that objects are encouraged to stay within the correct zones, 
\begin{equation}\label{eqn: in_bounds}
C_{\text{zone}, i} :=  \sum_{j \neq i}^k \min(\text{dist}(s_i, z_j) - \text{dist}(s_i, z_i), 0.0)^2. 
\end{equation}

The overall optimization takes the form, 
\begin{equation}
\begin{aligned}
 \min_{\{x_i, y_i, \theta_i\}_{Z_k}} C_{\text{sec}, k} := \lambda_3 C_{\text{over}}  +  \sum_{i \in Z_k} ( C_{\text{lang}, i} + \lambda_5 C_{\text{bound}, i} +\\ C_{\text{align}, i} +   C_{\text{wall}, i} + \lambda_6 C_{\text{zone}, i} ).
\end{aligned}
\end{equation}
Note, compared to \autoref{eqn:pri_place} we add a constraint for zoning here and remove $C_\text{bal}$. Once the secondary objects are fixed for a zone, we update the zone centroids by calculating the mean coordinates of all objects (primary and secondary) within that zone. We then redefine the zone boundaries using a new Voronoi segmentation based on the updated centroids.

After the secondary objects have been placed in all zones, we proceed to incorporate the tertiary objects.

\paragraph*{C. Tertiary object placement.}
For this step, we use an altered set of default constraints that ensures that objects of the same type cannot overlap, and that tertiary objects that should be wall-mounted are both on the wall ($\omega$) and that they are avoiding intersections with doors and windows.  

In the final stage of optimization, we find the location and orientation $(x_i, y_i, \theta_i)$ of all of the tertiary objects ($i \in \mathcal{T}$) at once, regardless of zone. We do these all at once since each object has only one constraint making the optimization simpler. In \autoref{eqn: tertiary} and \autoref{eqn: on_wall}, $C_{\text{over[i, j]}}$ is the cost only between objects $i$ and $j$, $\lambda_7$ and $\lambda_8$ are tunable parameters (we use 500 for both in our experiments), $I_{type_i = type_j}$ is an indicator variable that has value 1 if objects $i$ and $j$ have the same type, otherwise 0, and $I_\omega$ is an indicator variable that has value 1 if the object is wall-mounted, otherwise 0.
  
The optimization takes the form, 
\begin{equation}\label{eqn: tertiary}
\begin{aligned}
\min_{\{x_i, y_i, \theta_i\}_{\text{ter}}} C_{\text{ter}} :=  \sum_{i \in \text{ter}} [C_{\text{lang}, i} + \lambda_7 C_{\text{bound}, i} + C_{\text{align}, i}+ I_\omega C_{\text{on\_wall}, i} \\ + \sum_{j \in \text{ter}, j > i}(I_{type_i = type_j})C_{\text{over}[i, j]}]. 
\end{aligned}
\end{equation}
with the wall alignment cost being defined as, 
\begin{equation}\label{eqn: on_wall}
\begin{aligned}
   C_{\text{on\_wall}, i} := \sum_{j \in \text{doors} \cup \text{windows}} \lambda_8 C_{\text{over}[i, j]} \\ + \prod_{\omega \in \text{walls}} \left( \text{dist}(t_i, \omega) + (\theta_i - \theta_{\omega})^2 \right).
   \end{aligned}
\end{equation}

\subsection{Object Retrieval and Visualization}
Having generated the final layout, we retrieve the objects based on their generated descriptions and add them to the scene for visualization. %
For each object (including windows and doors), we search, using text, for an asset that matches the style description generated, as described before.
We scale the retrieved objects based on target width/depth, while proportionally scaling the height. We orient the objects based on the target angle $\theta_i$ assuming the objects have consistent (front) orientation. 
We source these assets using BlenderKit~\cite{BlenderKit}, and apply the same process for the wall and floor materials. In isolated cases, we manually modify the materials of the assets to better align with the descriptions produced by the LLM. (The only other manual adjustments made in this phase are for adding lighting for rendering.) 
We note that this phase can be better automated using  CLIP~\cite{clip} for object retrieval, leveraging text-image similarity scores to fetch objects from Objaverse~\cite{objaverse}, as employed in competing methods like Holodeck~\cite{holodeck}. Also, linking to a generative 3D modeling system will reduce the reliance on the models in the database -- this is left for future exploration.

\begin{figure*}[t!]
    \centering
\includegraphics[width=.89\linewidth]{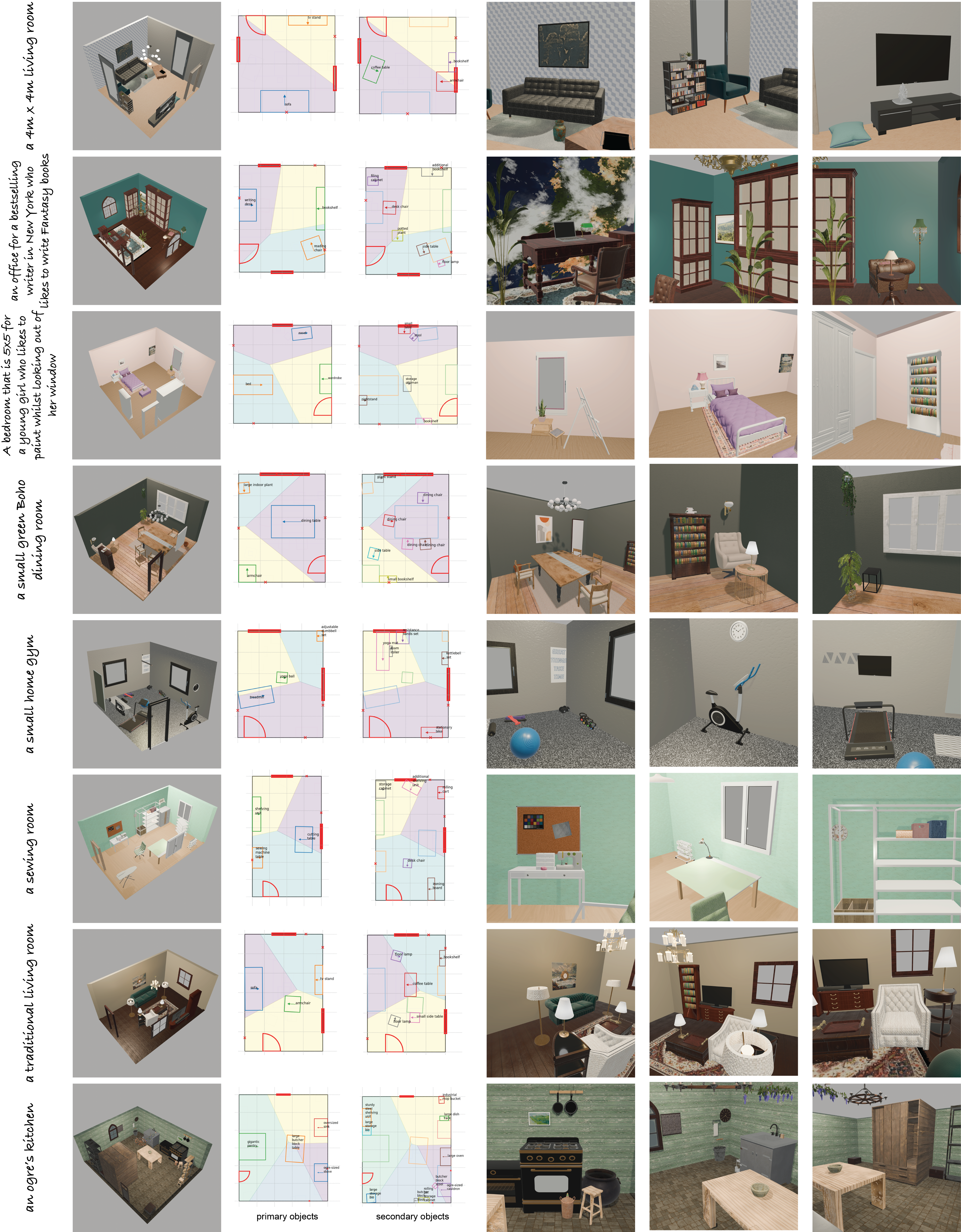}
    \caption{Results. We showcase a diverse range of layouts generated by \name, covering a wide range of prompts—from traditional bedroom and living room designs to more specialized spaces like a sewing room, as well as highly stylized concepts. From left to right, the visualizations include the text prompt, a three-quarter view, a floor plan highlighting primary objects, a floor plan detailing secondary objects (tertiary ones are not shown on floorplan), and close-up views for finer detail. See supplemental for walkthroughs. }
    \label{fig:results}
\end{figure*}

\section{Evaluation}
\label{sec:evaluation}
We compare our approach with four recent LLM-based methods, namely LayoutGPT~\cite{layoutgpt}, HoloDeck~\cite{holodeck}, and I-Design~\cite{idesign} as well as with transformer-based layout generator ATISS~\cite{ATISS}. We quantitatively evaluate the layouts on practical measures such as accessibility (pathway), area of overlapping objects, and area occupied by objects that are out of bounds. We also conduct a user study to qualitatively compare the quality of layouts and see how ours performs compared to layouts created by amateurs. We also conduct an ablation study to prove the effectiveness of our design choices. We note that since LayoutGPT, ATISS, and I-Design do not generate windows or doors in their scenes, we use the doors and windows generated by our method throughout.

\subsection{Metrics}\label{sec:metrics}
\begin{enumerate}[(i)]
    \item \textit{Pathway cost:} We design a cost function to evaluate the clearance of pathways in a room to measure accessibility/walkability. The pathway is generated using the medial axis of the room boundary and the floor objects (primary and secondary objects) and is then expanded to a width of 0.6 m. This pathway is represented as a set of points ($P$), and for each primary or secondary object, we check if any of these pathway points lie within their bounding box $B_i$. If a point is inside the bounding box, we compute the squared distance from the pathway point to the nearest object boundary ($\partial B_i$), as 
    \begin{equation}
        C_{\text{pathway}} := \sum_{i\in \{pri,sec\}} \sum_{p \in P \cap B_i} \left[ d\left( p, \partial B_i \right) \right]^2. 
    \end{equation}
    \item \textit{Object overlap rate (OOR):} In a good design layout, there should be no overlap between objects. We calculate the rate of overlapped objects as follows:
    \begin{equation}\label{eqn:oor}
    OOR := \frac{\sum_{i}\sum_{j>i}A_{\text{over}[i, j]} + \sum_{q} \sum_{r > q} A_{\text{over}}[q, r] (I_{type_q = type_r})}{w \cdot l} 
    \end{equation}
    where $A_{\text{over}}[i, j]$ is the area of overlap between objects $i$ and $j$ (including intersections with door buffers that account for the door swing area), $I_{type_q = type_r}$ is an indicator variable that has value 1 if tertiary objects $q$ and $r$ have the same type, otherwise 0; $w$ and $l$ are the width and the length of the room respectively.
    \item Out of Bounds Rate (OOB): All objects must fit fully inside a room for practicality. We measure the rate of area occupied by objects, which is out of bounds as follows:
    \begin{equation}\label{eqn:oob}
    OOB := \frac{\sum_{i}A_\text{bound}[i]}{w \cdot l}
    \end{equation}
    where $A_{\text{bound}[i]}$ is the area out of bounds for object $i$. 
\end{enumerate}

\begin{figure}[b!]
    \centering
\includegraphics[width=\columnwidth]{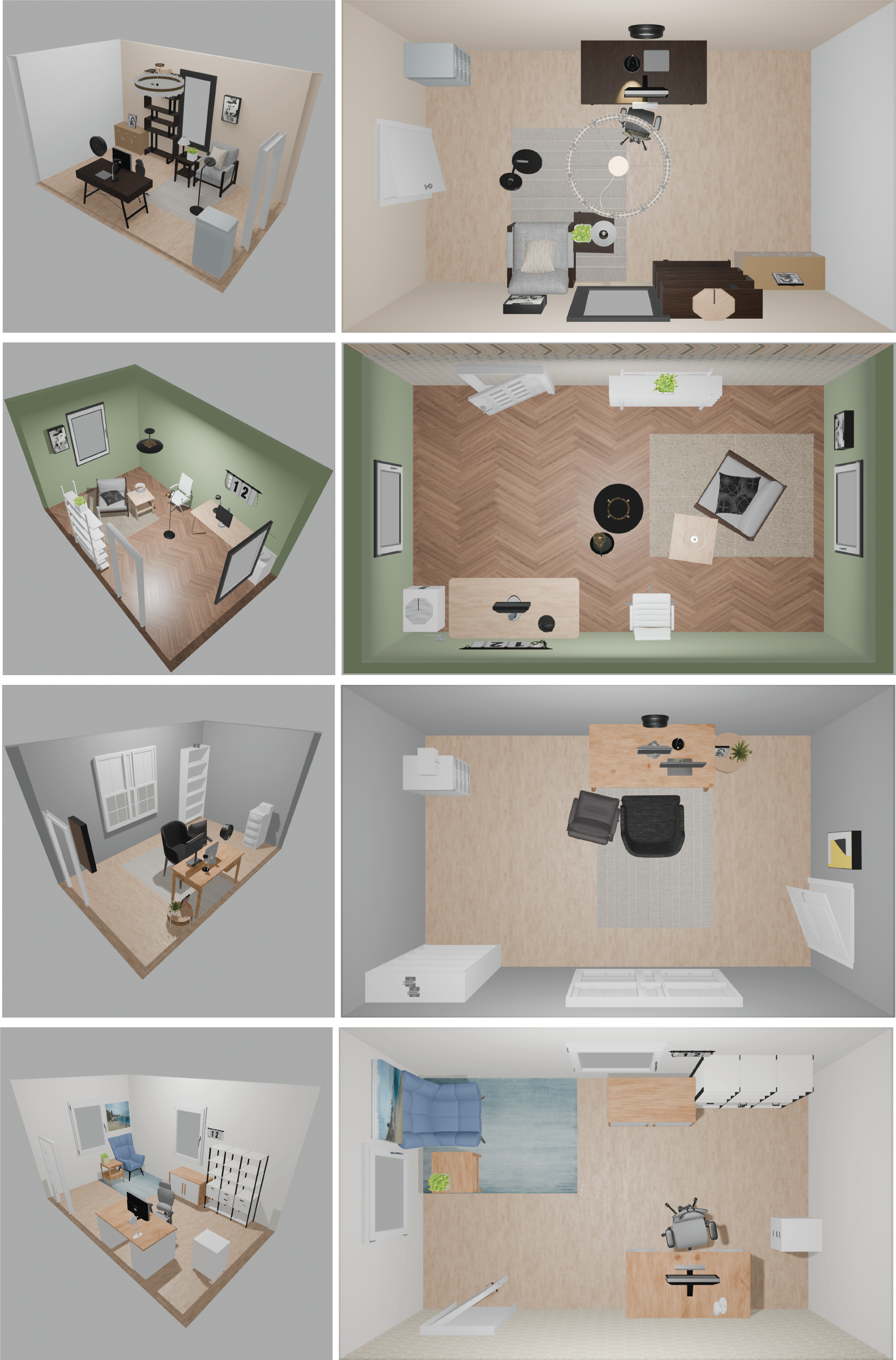}
    \caption{Diversity in Generated Layouts. \name demonstrates impressive versatility in scene generation for the same input prompt-``A 5m x 3m home office'', producing a wide range of layouts driven by variations in the selection of objects and style (guided by the LLM), and placement of windows, doors, and sockets. These elements significantly influence the arrangement of objects during our optimization phase, resulting in diverse and dynamic room configurations. We show primary and secondary objects on the left and on the right we show tertiary objects.}
    \label{fig:designDiversity}
\end{figure}

\subsection{Quantitative Evaluation} We compare our \name with both closed-universe and open-universe LLM-based layout generation methods—LayoutGPT~\cite{layoutgpt}, Holodeck~\cite{holodeck} and I-Design~\cite{idesign}. The comparison is based on the three metrics which measure practicality as outlined in \autoref{sec:metrics}, with results presented in \autoref{table:quant_comparison}. \name significantly outperforms LayoutGPT and Holodeck across all metrics, and I-Design for all but the out of bounds rate. LayoutGPT, as a closed-universe approach, is constrained to generating standard layouts for bedrooms and living rooms, lacking the flexibility to create more stylized or unique designs. Please note that our method does not explicitly add cost functions for pathway ($C_{\text{Pathway}}$) but we ensure walkability as a result of our wall-attraction cost, which encourages suitable objects
to be near the wall as well as our customizable accessibility constraints mapped by the LLM during the language phase. While running I-Design, we observed that in several instances, it was necessary to restart the process from the beginning, starting with the prompt. This was because the optimization phase often failed to converge, even after three hours. The strong quantitative results highlight the feasibility of our method compared to naseline methods. Comapred to other LLM-based + optimization-driven approaches, such as Holodeck’s constraint-based layout design and I-Design’s backtracking strategy, our cost based optimzation proves to be more effective at ensuring functionality.

\begin{table*}[h]\centering
\caption{Comparison.
Quantitative comparison against different methods measuring the functionality of the generated layouts in terms of object accessibility~(OOB), object overlap~(OOR), and access pathway~($C_{\text{Pathway}}$).}
\resizebox{\textwidth}{!}{%
  \begin{tabular}{r rrr rrr rrr rrr}
    \toprule
    \multirow{2}{*}[-0.5\dimexpr \aboverulesep + \belowrulesep + \cmidrulewidth]{Prompt}
    & \multicolumn{3}{c}{\textbf{LayoutGPT}~\cite{layoutgpt}} & \multicolumn{3}{c}{\textbf{Holodeck}~\cite{holodeck}} & \multicolumn{3}{c}{\textbf{I-Design}}~\cite{idesign}& \multicolumn{3}{c}{\textbf{\name (ours)}} \\
    \cmidrule(l){2-4} \cmidrule(l){5-7}\cmidrule(l){8-10}\cmidrule(l){11-13}
    & OOB $\downarrow$ & OOR $\downarrow$ & $C_{\text{Pathway}}$ & OOB $\downarrow$ & OOR $\downarrow$ & $C_{\text{Pathway}}$ & OOB $\downarrow$ & OOR $\downarrow$ & $C_{\text{Pathway}}$ & OOB $\downarrow$ & OOR $\downarrow$ & $C_{\text{Pathway}}$\\
    \midrule
    ``A bedroom that is 3m x 4m.'' & 0.773 & 3.973 & 	12.315 & 0.890 & 0.332 & 3.764 & 0.000 & 5.540 & 3.943 &  0.095 & 0.000 & 0.291  \\
    ``A bedroom that is 3.225m x 4.5m.'' & 4.752 & 0.000 & 12.617 & 1.630 & 1.532 & 1.163 &  0.000 & 3.148 & 5.302 &0.215 &  0.004 & 2.916\\
    ``A bedroom that is 4.3m x 6m.'' & 2.920 & 3.518 & 4.173 & 1.518 & 0.000 &	2.828 & 0.000  & 5.860 & 2.166 & 0.009 &	0.008 & 1.406 \\
    ``A bedroom that is 5m x 5m.'' & 0.000 & 0.811 & 	10.569 & 2.013 & 1.242 & 5.129 & 0.000 & 4.096& 1.224 & 0.010 & 0.012 & 0.000 \\
    ``A bedroom that is 3m x 8m.'' & 1.129 & 10.080 & 1.843 & 1.412 & 0.000 &	5.650 & 0.000 & 2.046 & 4.201 &0.005 &	0.003 & 3.678\\
    ``A living room that is 5m x 5m.'' & 2.040 &	6.480 &	2.958 & 0.996 & 0.000 & 6.240 & 0.000 & 7.166 & 5.028 &0.000 &	0.004 &	0.740 \\
    ``A living room that is 3m x 4m.'' & 0.001 &	7.046 &	2.010 & 2.013 &	2.200 & 6.712 & 0.000 & 3.854 & 1.697 &0.074 & 0.000 & 0.204\\
    ``A living room that is 4m x 6m.'' & 4.427 &	1.282 & 0.852 & 1.611 & 3.215 & 8.021 & 0.000 & 3.700 & 6.284 &0.019 & 0.008 & 0.050\\
    ``A living/dining room that is 6m x 3m.'' & 7.978 &	7.582 &	3.092 & 2.191 &	0.000 &	0.605 & 0.000 & 6.582 & 2.169 &0.061 &	0.007 &	5.154 \\
    ``A living room that is 8m x 4m.'' & 0.000 &	5.488 &	10.843 & 1.022 & 0.079 & 17.479 & 0.000 & 0.147 & 7.088 &0.048 & 0.030 & 1.027\\
    ``A bedroom that is 4m x 5m.''& 2.219 & 9.138 & 8.993 &	1.840 &	1.949 & 6.463 & 0.000 & 1.353 & 2.719 &0.007 & 0.017 & 0.735 \\
    ``A sewing room.'' & \ding{55} & \ding{55} & \ding{55} & 1.317 &0.000 &	10.699 & 0.000 & 4.030 & 0.456 &0.007 & 	0.000 & 1.033 \\
    ``A small green boho dining room.''  & \ding{55} & \ding{55} & \ding{55} & 1.971&	2.150 &	10.674 &  0.000 & 1.907 & 0.047 &0.100 & 0.011 & 1.639 \\
    \multirow{ 2}{5.4cm}{``An office for a bestselling writer in New York who likes to write Fantasy books.''} & \multirow{ 2}{*}{\ding{55}} & \multirow{ 2}{*}{\ding{55}} & \multirow{ 2}{*}{\ding{55}} & \multirow{ 2}{*}{1.659} & \multirow{ 2}{*}{0.365} & \multirow{ 2}{*}{6.176} & \multirow{ 2}{*}{0.000} &	\multirow{ 2}{*}{2.513} &	\multirow{ 2}{*}{5.572}& \multirow{ 2}{*}{0.010} &	\multirow{ 2}{*}{2.588} &	\multirow{ 2}{*}{2.262} \\
     &  & &  &  & &	 &  & & & & & \\
    ``A bedroom for a vampire.'' &\ding{55} & \ding{55} & \ding{55} &  1.683 & 0.302 &	3.982 & 0.000  & 9.710 & 2.048 &0.043 & 0.094 & 2.469 \\
    \midrule 
    Mean Scores & 2.385 & 5.036 & 6.388& 1.584 & 0.891 & 6.372 & \textbf{0.000} & 4.110 & 3.330 & 0.047 & \textbf{0.186} & \textbf{1.736}\\
    \bottomrule
  \end{tabular}}
  \label{table:quant_comparison}
\end{table*}

\subsection{Qualitative Evaluation} We present the results of our method in \autoref{fig:results}, showcasing layouts generated from a variety of prompts. These range from traditional bedroom and living room designs to more specialized spaces, such as a sewing room, and stylized concepts like ``A small workroom for a wizard.''  \name also demonstrates its ability to meet specific client-driven functional and aesthetic requirements, such as ``A bedroom that is 5x5 for a young girl who likes to paint whilst looking out of her window'' or ``An office for a bestselling writer in New York who likes to write Fantasy books''. 

\begin{figure}[h]
    \centering
\includegraphics[width=\columnwidth]{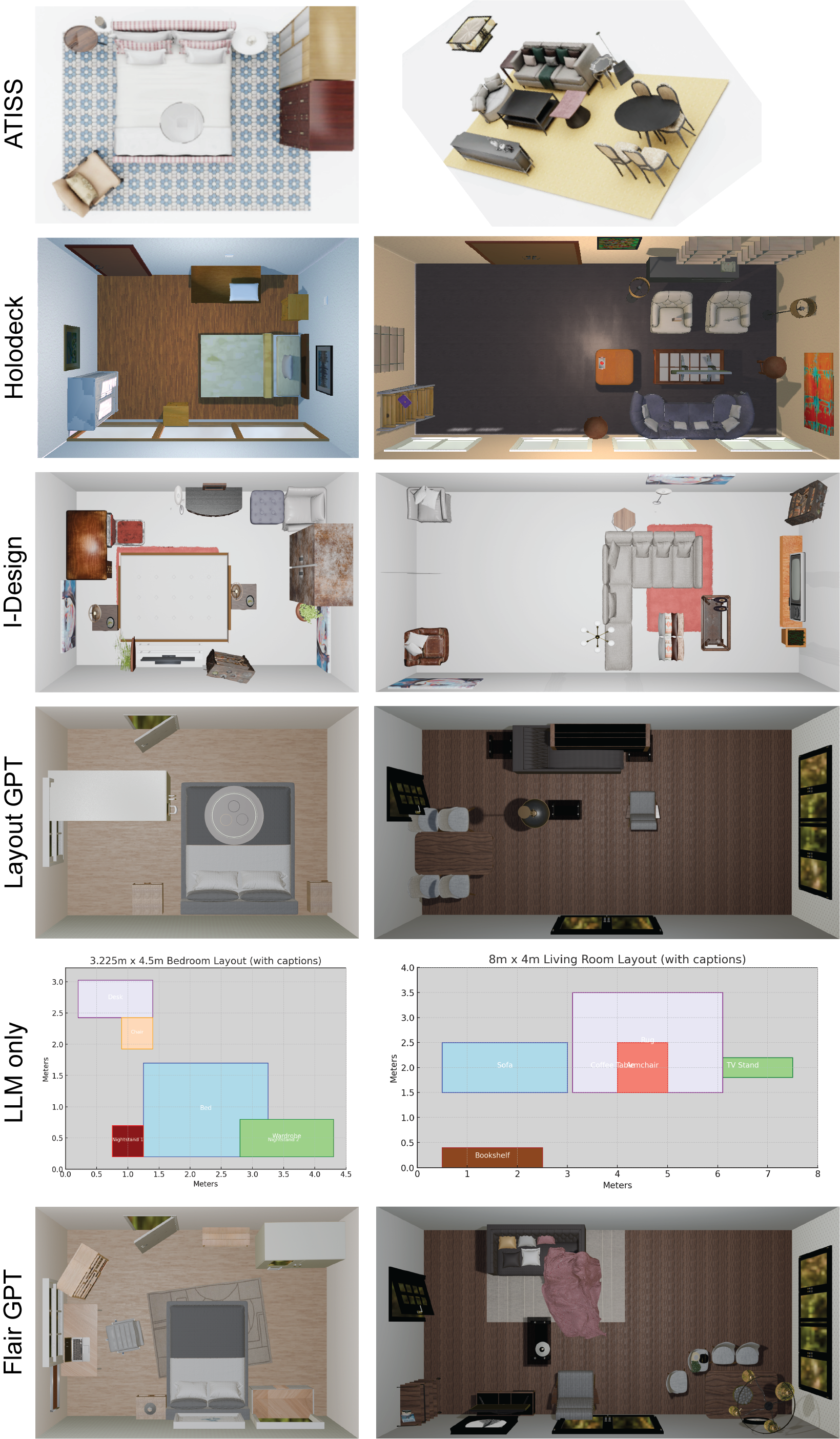}
    \caption{Room layout comparison against baselines. Comparison of layouts generated by \name and baseline methods, highlighting differences in object arrangement, spatial organization, and overall design quality.}
    \label{fig:comparison}
\end{figure}

We compare our method against baseline approaches—LayoutGPT~\cite{layoutgpt}, I-Design~\cite{idesign} and Holodeck~\cite{holodeck}—in \autoref{fig:comparison} . Our results demonstrate a closer alignment with the input prompt for stylized designs. For instance, in the prompt ``A bedroom for a vampire,'' the generated layout replaces the traditional bed with a coffin, showcasing \name's creative and context-aware object selection to match the thematic style of user prompts. Video results are available on the \href{https://flairgpt.github.io/}{supplemental webpage}. Additionally, \name can generate multiple distinct layouts for the same input prompt, as seen in \autoref{fig:designDiversity}, offering versatility and a range of design options that cater to individual preferences and specific requirements. Please refer to our supplemental \href{https://flairgpt.github.io/}{webpage} for a thorough comparison against baseline methods on nine varied layouts. These results demonstrate \name's efficient and comprehensive use of the layout, outperforming baseline methods that either create clutter or leave excessive empty space.

\paragraph*{User Study I.}
In this study, we asked users to compare \name against five methods: the first  four approaches are computational (LayoutGPT~\cite{layoutgpt}, ATISS~\cite{ATISS}, Holodeck~\cite{holodeck}, and iDesign~\cite{idesign}), the  fifth one being novice human designers. We were unable to run ATISS directly as the model weights are not publicly available, so we used the results reported in their paper instead.

\begin{figure}[b!]
    \centering
\includegraphics[width=\columnwidth]{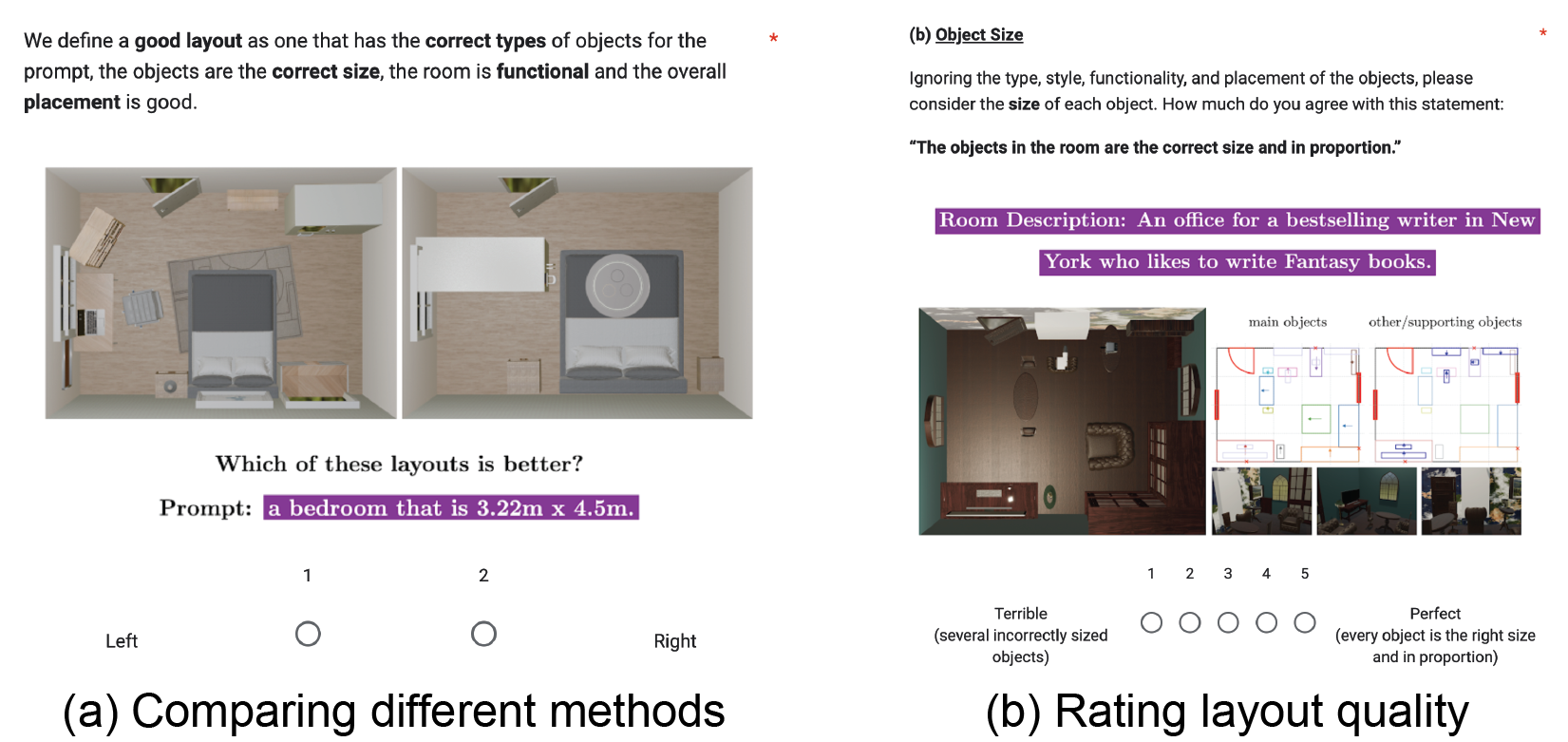}
    \caption{Screen Capture of User Studies. In User Study I (Figure (a)), participants compared \name with LayoutGPT, ATISS, and novice designers. In User Study II (Figure (b)), participants rated layouts by \name and novice designers across multiple criteria.}
    \label{fig:user_study}
\end{figure}

To compare our method against novice human designers, we recruited 5 participants to design 2 layouts each. Each participant was provided with two blank floorplans containing windows and doors positioned identically to those in our method (see supplemental for details). They had 15 minutes per floorplan to draw bounding boxes for each object in the room (along with forward direction), without guidance on object sizing. Participants had control over object selection, size, orientation, and placement, but not style. From these designs, we selected 4 layouts (2 for each prompt) and reconstructed them in Blender using the same assets as our method. If a participant included objects that were not present in our room inventory, we selected assets that matched the specified style.

For the closed-universe computational methods, we study three different simple prompts, for the open-universe computational methods, one simple prompt and two more complex prompts, and for the human method, two:
\begin{itemize}
\item Closed-Universe Computational:
\begin{enumerate}[(i)]
    \item ``A bedroom that is 3m x 4m.''
    \item ``A bedroom that is 3.225 x 4.5m.''
    \item ``A living room that is 8m x 4m.''
\end{enumerate}
\item Open-Universe Computational:
\begin{enumerate}[(i)]
    \item ``A bedroom that is 3m x 4m.''
    \item ``A sewing room.''
    \item ``A bedroom that is 5m x 5m for a young girl who likes to paint whilst looking out of her window.''
\end{enumerate}
\item Human:
\begin{enumerate}[(i)]
  \setcounter{enumi}{3}
  \item ``A bedroom that is 4m x 5m.''
  \item ``An office for a bestselling writer in New York who likes to write Fantasy books.''
\end{enumerate}
\end{itemize}

Participants were shown bird’s-eye renderings of each method and condition, similar to \autoref{fig:user_study} (a). In an unlimited-time, two-alternative forced choice task, they were asked to choose the ``better layout'' based on aesthetics, functionality, and adherence to the prompt. A total of 21 participants participated in this experiment, with the outcomes presented in \autoref{table: user_study}. 

We see that subjects prefer our results on average across prompts in 88.9\% of the cases over LayoutGPT, in 79.4\% of the cases over ATISS, in 81.7\% of the cases over Holodeck, in 95.0\% of cases over I-Design, and in 63.2\% of the cases over a human result (significant, $p<10^{-6}$, binomial test). Similar conclusions can be drawn when looking at individual prompt conditions (significant, $p<0.01$, binomial test).

\paragraph*{User Study II.}

Study 2 uses the same methods and similar viewing conditions as study 1, using the same prompts for the human baseline, but 5 prompts for our method: 

\begin{enumerate}[P1.]
\item ``A bedroom that is 4m x 5m.''
\item ``An office for a bestselling writer in New York who likes to write Fantasy books.''
\item ``A sewing room.''
\item ``A small green boho dining room.'' 
\item ``A bedroom for a vampire.'' 
\end{enumerate}

\begin{figure}[t!]
    \centering
\includegraphics[width=0.7\columnwidth]{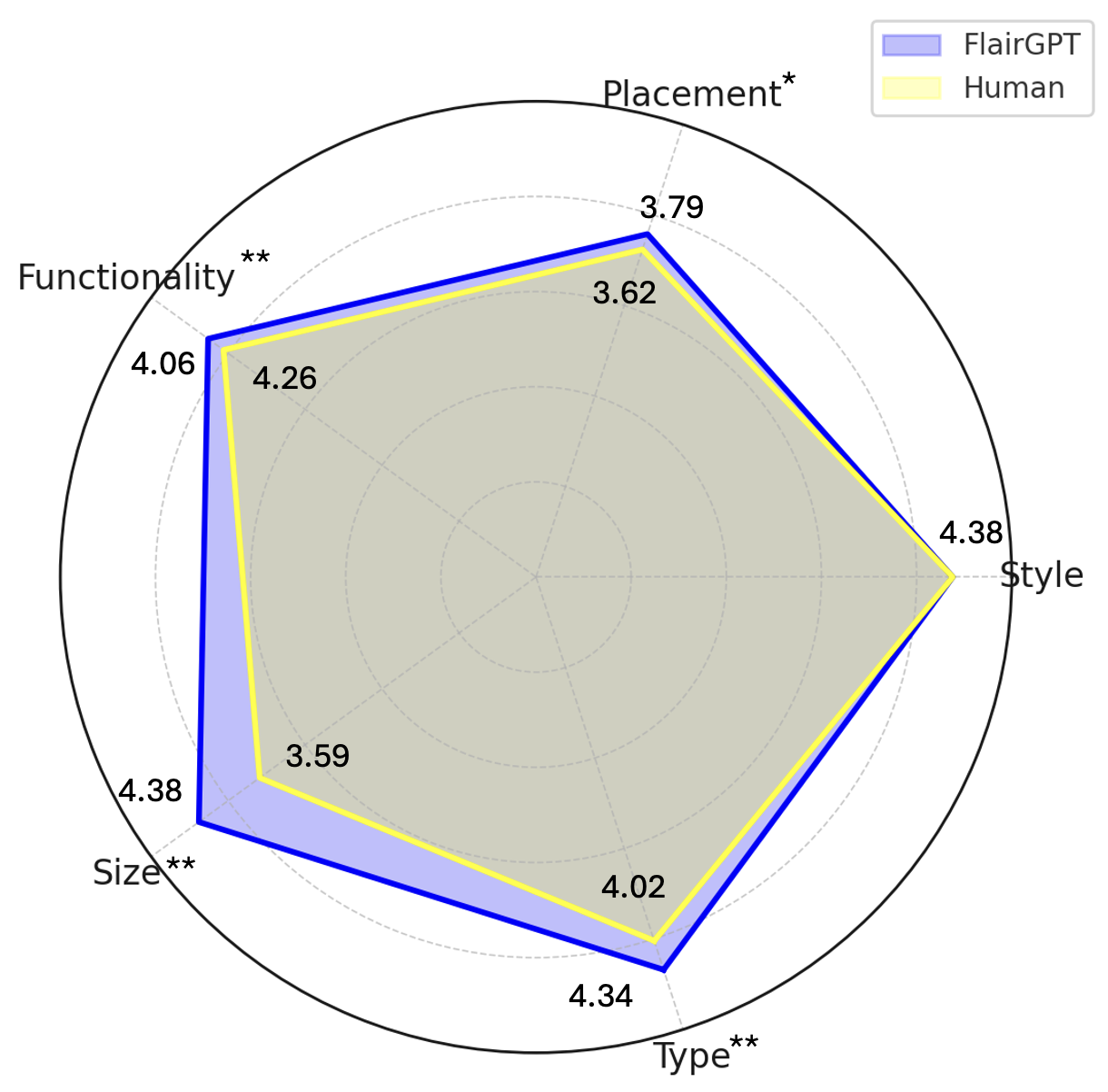}
    \caption{User Study II: Score comparison between \name and layouts designed by novices. Mean scores (out of 5) are shown for object type, object size, object style, object functionality and overall placement.  Each criterion was rated on a scale from 1 (terrible) to 5 (perfect). Since we used the assets chosen by our method for the human designed layouts, we use our score for both \name and the human designs. For three criteria, the difference was significant at $p < 0.001$ (**) and for one, it was significant at $p < 0.01$ (*).}
    \label{fig:user-criteria}
\end{figure}

Participants were shown a single result of a single method (as can be seen in \autoref{fig:user_study} (b)) and asked to rate with unlimited time on a five-point Likert scale according to five criteria: ``object type'', ``object size'', ``object style'', ``object functionality'', and ``overall placement''. We compare the four layouts drawn by novice human designers against the same four prompts picked from our generated results. A total of 17 participants participated in this experiment, where \name performed well across all criteria, as shown in \autoref{fig:user-criteria}. For the direct comparison between our layouts and the human-designed ones, we excluded the style criterion since the rooms were constructed using the style produced by our method. Participants rated our method, aggregated across four criteria and all rooms, at 4.19 compared to 3.82 for human designs (difference significant at $p<0.0001$, $t$-test).

\begin{table}
\begin{minipage}{\columnwidth}
\centering
\caption{User Study Findings.
Users preferred layouts generated by \name over those by LayoutGPT, ATISS, Holodeck, or I-Design. When compared against human designers, ours was preferred for more complex/creative prompts~(P5), while human designers were better in the simple/standard scenario~(P4). \\
\\(a) \name vs LayoutGPT, ATISS, Holodeck and I-Design across three prompts.}
\begin{tabular}{rcccc}
    \toprule
    Prompt & P1 & P2 & P3 & Average\\
    \midrule
    vs LayoutGPT & 85.7\% & 100\% & 81.0\% & 88.9\% \\
    vs ATISS & 81.0\% & 100\% & 57.1\% & 79.4\% \\
    \bottomrule
\end{tabular}

\vspace{0.5cm}

\begin{tabular}{rcccc}
    \toprule
    Prompt & P1 & P4 & P5 & Average\\
    \midrule
    vs Holodeck & 80.0\% & 100\% & 65.0\% & 81.7 \%\\
    vs I-Design & 100\% & 90.0\% & 95.0\% & 95.0\%\\
    \bottomrule
\end{tabular}
\label{table: user_study}

\vspace{0.5cm}

\textit{(b) \name vs novice human designers across two prompts.}

\vspace{0.5cm}

\begin{tabular}{rccc}
    \toprule
    Prompt & P6 & P7  & Average\\
    \midrule
    vs Human & 29.4\% & 94.1\% & 63.2\% \\
    \bottomrule
\end{tabular}
\end{minipage}
\end{table}

\paragraph*{LLM-based assessment.} 

In our research, we aimed to test the ability of LLMs to evaluate the quality of a layout. Specifically, we sought to determine whether an LLM could classify a layout as “good” or “bad” and identify potential flaws in the design. To explore this, we conducted an experiment with 24 bedroom layouts, some intentionally flawed and others well-designed. Four human participants labeled each layout as either “good” or “bad” and provided reasoning for their classifications.

We extended this evaluation to both GPT-4o and SigLIP \cite{siglip} using the same set of layouts. For this, we created four representations of each bedroom: a bounding box representation, a top-down 2D view, a top-down 3D view, and a perspective view from an angle chosen (for best visibility) within the 3D room. Each representation was individually presented to GPT-4o, which was tasked with listing the pros and cons of the layout before classifying it as either good or bad.

For SigLIP, we employed the same bedroom representations, pairing each with three captions: a positive caption (“a good layout for a bedroom”), a neutral caption (“a layout for a bedroom”), and a negative caption (“a bad layout for a bedroom”). We calculated similarity scores between the captions, denoted as $G_i$ for good, $N_i$ for neutral, and $B_i$ for bad, and the images. A layout was classified as good if $2G_i - N_i - B_i > 0$.

Our findings revealed that both GPT-4o and SigLIP performed best when using the 3D top-down view of the room. However, the accuracy of correct classifications was insufficient for practical use, with GPT-4o achieving 63\%. %

\subsection{Ablation}
We ablate our choice of cost constraints- $C_\text{bound}$ and $C_{\text{over}}$ as well as our hierarchical structure and cleaning step in \autoref{table: ablation}. Specifically, we compare our method without the boundary cost ($C_{\text{bound}}$), without the overlap cost ($C_{\text{over}}$), without the constraint cleaning phase, and with all objects optimized simultaneously rather than following our proposed hierarchical structure (for this, we allowed the optimization to run for 1.5 hours before taking the best result; for comparison, ours takes ~10-15 minutes on average). We evaluate these variants using the same out of bounds (OOB) and object overlap rate (OOR) as described earlier. We also measure translation errors (TE) which is described as $\frac{\text{number of translation errors}}{\text{number of uncleaned constraints}}$.

\begin{table}[h]\centering
\caption{Ablation. 
Our ablation results underscore the critical role of the additional cost constraints, our hierarchical optimization structure, and the cleaning step in enhancing the overall performance of our method.}
  \begin{tabular}{rcccc}
    \toprule
    Method & OOB $\downarrow$ & OOR $\downarrow$ & TE$\downarrow$  \\
    \midrule
w/o $C_{\text{bound}}$ & 9.20 & 0.01 & 15.70 \\
w/o $C_{\text{over}}$ & 0.03 & 3.68 & 15.70\\
w/o Hierarchy & 8.84 & 2.18 & 15.70 \\
w/o Cleaning &  0.04 & 0.23 & 19.24 \\
\name & 0.03 & 0.54 & 15.70\\
\bottomrule
\end{tabular}
\label{table: ablation}
\end{table}

\vspace{.1in}

\section{Conclusion}
\label{sec:conclusion}

We have presented \name as an LLM-guided interior designer. We demonstrated that LLMs offer a rich source of information that can be harnessed to help decide which objects to include for a target room along with their various intra- and inter-object constraints. We described how to convert these language constraints into algebraic functions using a library of pre-authored cost functions. Having translated the functions, we solve and extract final room layouts, and retrieve objects based on the LLM-based object attributes. Our evaluations demonstrate that human users favorably rate our designed layouts. The generated layouts are explainable by construction, as users can browse through the constraints used in the design process and optionally adjust their relative priority. Our approach, which leverages LLMs for planning followed by optimization, has the potential for wide-ranging applications beyond interior design, including urban planning, 3D modeling, and image enhancement--any domain where cost objectives can be effectively quantified.

\paragraph*{Limitations.} Our study has several limitations that future work could address. First, \name designs are currently limited to rectangular rooms. Exploring  application to irregularly shaped rooms, possibly by approximating them with union of  (axis-aligned) rectangles, would be an interesting direction. However, one has to come up with a canonical naming convention for the walls to interact with the LLM to extract room-specific constraints. 

Second, we pre-authored a set of cost functions for translating the LLM-specified constraints. In future work, we would like to investigate LLMs' generative capabilities to propose new cost functions for the library. Currently, we find that the algebraic reasoning skills of LLMs are inconsistent, making it challenging to develop an automated library generation capability. It is worth noting that our approach was zero-shot, as we did not fine-tune the LLM with example library functions.

Third, the object attributes do not have height associated with them, making it challenging to enforce constraints that prevent wall-mounted items from being placed behind taller objects — for example, a painting behind a wardrobe.

Finally, as described, we leave it to the LLM to decide and handle conflicting constraints in the constraint cleanup stage. Also, we fix the object size early in the pipeline when the LLM lists the room objects -- this restricts possible adjustments in the subsequent optimization phase. In the future, when LLMs can quantitatively evaluate layouts, or their descriptions, then one can imagine an outer loop to backpropagate errors to update the list of selected objects and/or their relevant constraints, and decide which objects or constraints to drop. 

\paragraph*{Acknowledgments.} 
We thank Rishabh Kabra, Romy Williamson, and Tobias Ritschel for their comments and suggestions. NM was supported by Marie Skłodowska-Curie grant agreement No.~956585, gifts from Adobe, and the UCL AI Centre.

\bibliographystyle{eg-alpha-doi}  
\bibliography{main_roomGenerator}

\onecolumn
{\centering{\Large{\textbf{Supplementary Material for \name: Repurposing LLMs for Interior Designs}}}}

\section*{Contents}
\begin{enumerate}
\item Statistics For Experiments (page 1)
\item Token Cost Comparison with I-Design~\cite{idesign} and Holodeck~\cite{holodeck}. (page 2)
\item User Study I Responses (page 3)
\item User Study II Responses (page 4) 
\item Human Forms for User Studies and Human Drawn Layouts (page 6)
\item Blank Constraint Cost Functions (page 9)
\item Full example language output for ``a bedroom that is 4m $\times$ 5m.'' (page 17)

\end{enumerate}

\newpage

\section{Statistics For Experiments}

\begin{table*}[htbp]\centering
\caption{Statistics for our experiments including: the number of primary (P), secondary (S), and tertiary (T) objects per scene; the number of constraints before cleaning, after cleaning, and after translation (function calls); the number of errors including Language errors, Cleaning errors, Translation errors, and Optimization errors; and the time (minutes) for the Language and Translation phase combined, the Optimization phase, and the total time to generate each layout. }

\resizebox{\textwidth}{!}{  
\begin{tabular}{lcccccccccccccc}
    \toprule
    \multirow{2}{*}[-0.5\dimexpr \aboverulesep + \belowrulesep + \cmidrulewidth]{Prompt}
    & \multicolumn{3}{c}{Objects} & \multicolumn{3}{c}{Constraints} & \multicolumn{5}{c}{Errors}&\multicolumn{3}{c}{Time (mins)} \\
    \cmidrule(l){2-4} \cmidrule(l){5-7}\cmidrule(l){8-12}\cmidrule(l){13-15}
    &  P & S & T & Uncleaned & Cleaned & Function Calls & Language & Cleaning & Translation & Contradiction & Optimization & Language + Translation & Optimization & Total\\
    \midrule
"A bedroom that is 4m x 5m." &  3 & 4 & 7 & 49 & 52 & 57 & 1 & 2 & 6 & 0 & 1 & 0.82 & 7.20 & 8.02\\
"A living room that is 4m x 4m." & 2 & 3 & 10 & 43& 45& 48&  1& 2 & 7 &1  & 1 & 1.16 & 7.60 & 8.76 \\
"A sewing room." &  3 &  5 & 11 & 59 & 62 & 70 & 0 & 1 & 11 & 2 & 1 & 1.06 & 12.71 & 13.76 \\
"A small home gym." &  3 &  5 & 7 & 52 & 48 & 53 &   1& 0 &  6& 1 & 0 & 1.56 & 14.45 & 16.01\\
"A small green boho dining room." & 3 & 7 & 9 & 58 & 65 & 68 & 1 & 1 & 24 & 2 & 0 & 1.05 & 24.35 & 25.41 \\
"A traditional living room." & 3 & 5 & 10 & 64 & 73 &  72 &  0 & 2 &  7&  1 & 3 & 1.37 & 8.17 & 9.54 \\
\multirow{3}{4.5cm}{"An office for a bestselling writer in New York who likes to write Fantasy books."} & & & & & & & & & & & & & &   \\
& 3 & 4 & 11 & 60 & 62 & 63 & 1 & 3 & 4 & 0 & 1 & 1.08 & 13.04 &  14.12 \\
& & & & & & & & & & & & & &   \\
\multirow{3}{4.5cm}{"A bedroom that is 5x5 for a young girl who likes to paint whilst looking out of her window."} & & & & & & & & & & & & & &   \\
&  3 & 5 & 8 & 62 & 62 & 61 & 0 & 3  & 16  & 1 & 1 & 1.03 & 6.97 & 8.00 \\
& & & & & & & & & & & & & &   \\
"A bedroom for a vampire." & 3 & 4  & 9 & 49 & 47 & 47 & 0 & 0 & 0 & 2 & 2 & 0.85 & 6.68 & 7.53 \\
"A small workroom for a wizard." & 3 & 6 & 10 & 65 & 64 & 65  & 0 & 0 & 6 & 1 & 1 & 1.24 & 10.85 & 12.08  \\
"A kitchen for an ogre." & 4 & 10 & 10 & 72 & 73 & 79 & 0 & 7 & 3 & 2 & 1 & 1.61 & 12.13 & 13.73  \\
\midrule
\textbf{Mean values} & \textbf{3.00} & \textbf{5.27} & \textbf{9.27} & \textbf{57.55} & \textbf{59.36} & \textbf{62.09} & \textbf{0.36} & \textbf{1.91} & \textbf{8.27} & \textbf{1.18} & \textbf{1.09} & \textbf{1.17} & \textbf{11.29} & \textbf{12.45} \\
    \bottomrule
  \end{tabular}}
  \label{table: stats}
\end{table*}

We define 5 types of errors that can occur throughout our method:

\begin{itemize}
    
\item Language Error: This type of error arises purely from the output of the LLM during the language generation phase. It includes incorrect object sizing, nonsensical constraints (e.g., “put the table lamp on the armchair”), or other errors in the initial LLM output.
\item Cleaning Error: These errors occur during the cleaning phase. Examples include the unintended removal of constraints or the omission of crucial information from a constraint.
\item Translation Error: This is the broadest category of errors and can occur at any point during the translation phase. It may involve matching a language constraint to a similar but suboptimal constraint (e.g., selecting “away from window” instead of “not blocking a window”), completely misinterpreting the constraint, missing applicable constraints that have matching functions, or using incorrect parameters. Translation errors are the most frequent type of error.
\item Contradictory Constraint Error: This error occurs when two or more constraints are chosen that are mutually exclusive, making it impossible to satisfy all of them simultaneously within the solution.
\item Optimization Error: An optimization error arises when an object is placed in a position that does not align with its constraints, and yet the optimization process fails to find a better solution throughout the optimization process.
\end{itemize}

While there are many places for errors to arise, they are not all critical. For example, the most common translation error that we have seen is choosing “ind\_away\_from” instead of “ind\_not\_block” which are similar constraints and will achieve the object not blocking the window. When incorrect types of parameters are used, the function returns 0 so that constraint is lost. This can occur when choosing the sides of an object (one of “left”, “right”, “front” or “back”) with the LLM choosing something like “longer side”. The most problematic errors are the contradictory constraint errors and the optimization errors. These are the most visible in the outputs, however these are also far less frequent than translation errors.

\subsection{Success Rate with Varying Numbers of Objects}

In our experiments, we found that the success was similar for a room with 5 primary and secondary objects (``A living room that is 4m x 4m") and a room with 14 objects (``A kitchen for an ogre"). The tertiary objects are optimized separately and rely on a maximum of one primary or secondary object, and so larger numbers should not interfere with the success too much. Our method was designed in such a way that there is not too much variation in the number of objects generated to prevent overcrowding. 

\vspace{0.3in}

\section{Token Cost Comparison with I-Design~\cite{idesign} and Holodeck~\cite{holodeck}}

\begin{tabular}{rccc}
    \toprule
    Method & Number of API Calls  &  Total Number of Tokens \\
    \midrule
    I-Design~\cite{idesign} & 23 $\pm$ 4.58 	 & 35295 $\pm$ 5355 \\
    Holodeck~\cite{holodeck} & 7 &   6440 $\pm$ 2007\\
    \name (Ours)  & 20.66 $\pm$ 1.53 &   22762 $\pm$ 2179
\\
    \bottomrule
\end{tabular}

\onecolumn

\section{User Study 1 Responses}
\begin{minipage}{\textwidth}
    \includegraphics[page=1, scale = 0.75]{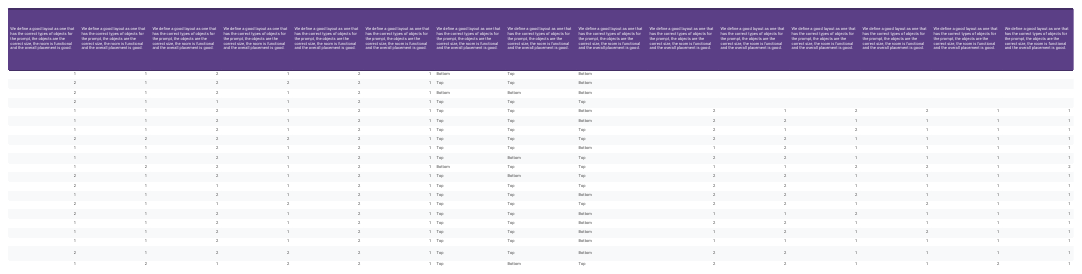}
\end{minipage}

\newpage
\section{User Study 2 Responses}
\begin{minipage}{\textwidth}
    \includegraphics[page=1, scale = 0.75]{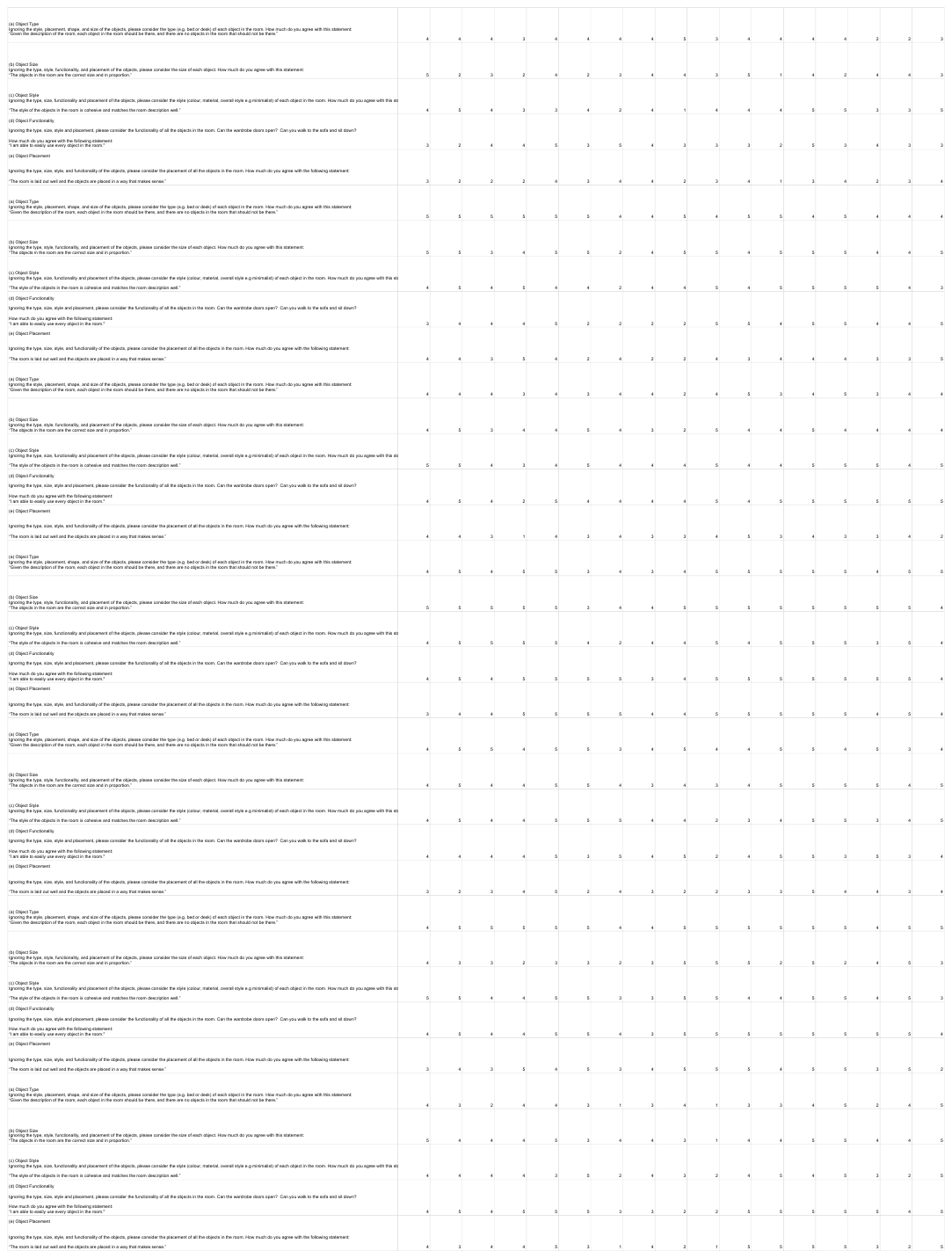}
\end{minipage}

\includepdf[pages=2, scale = 0.8]{supplementary/us2.pdf}

\section{Human Forms for User Studies and Human Drawn Layouts}
\begin{figure}[htbp]
    \centering
    \includegraphics[width = 0.5\textwidth]{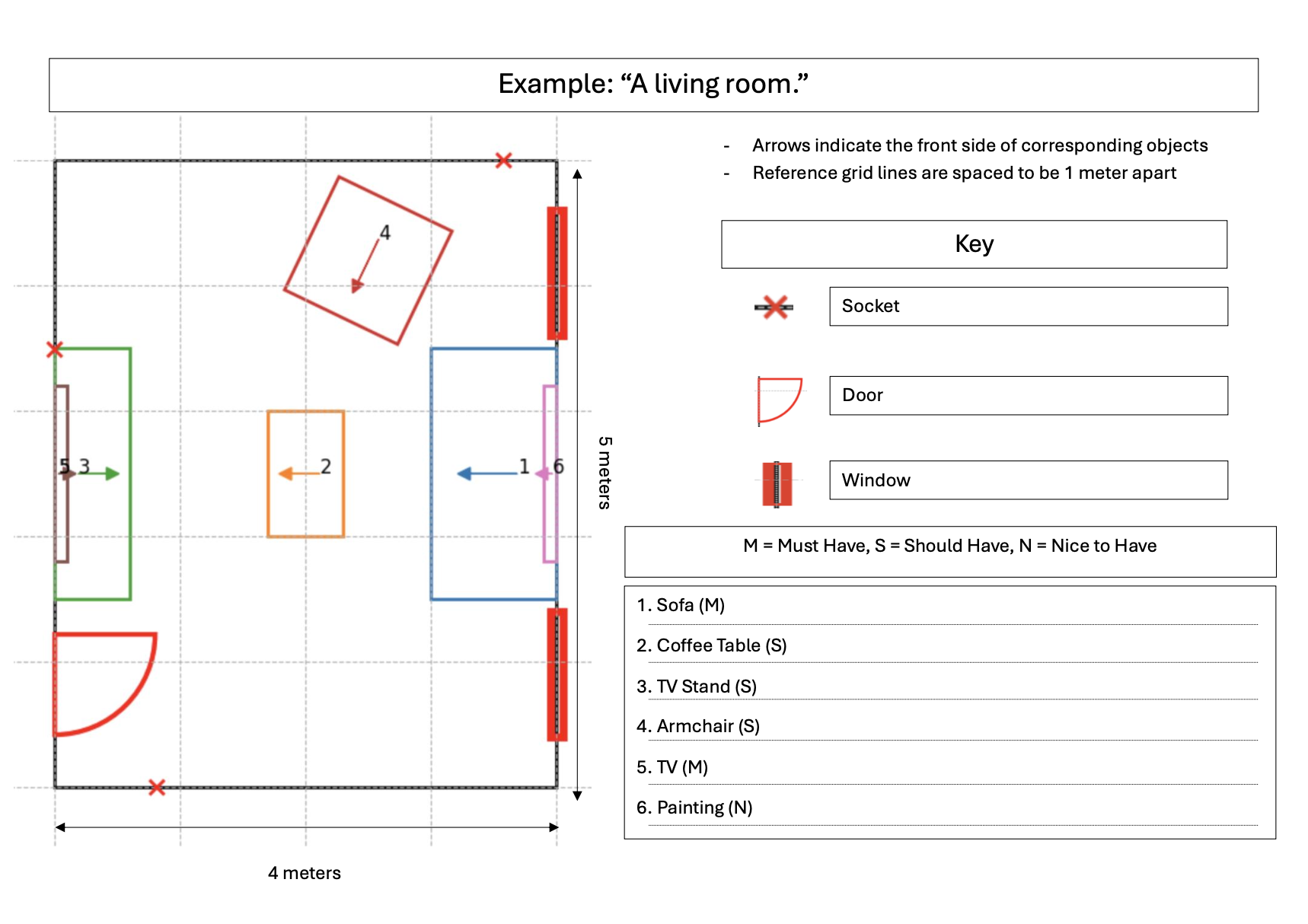}\\
    \includegraphics[width = 0.5\textwidth]{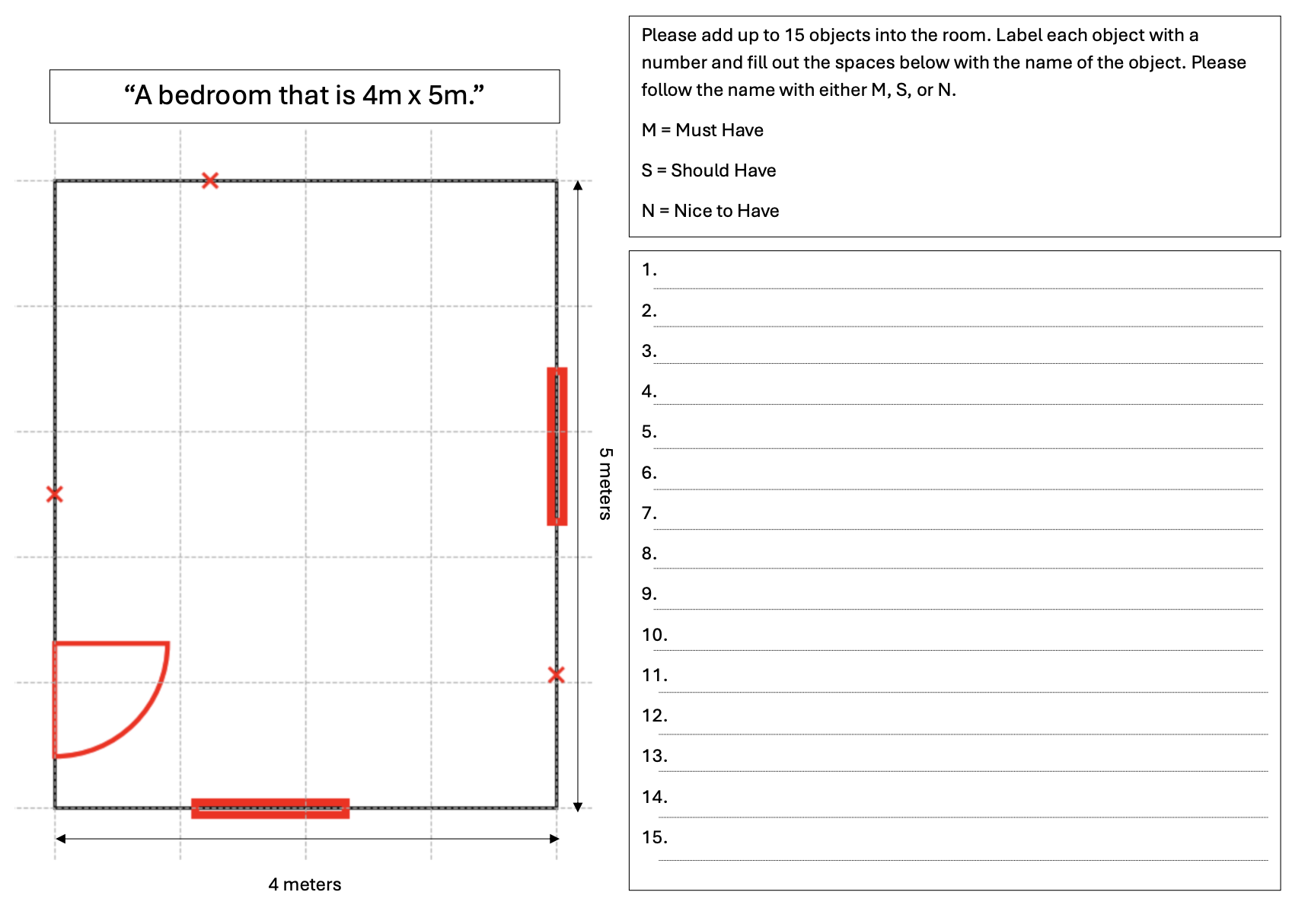}\\
    \includegraphics[width = 0.5\textwidth]{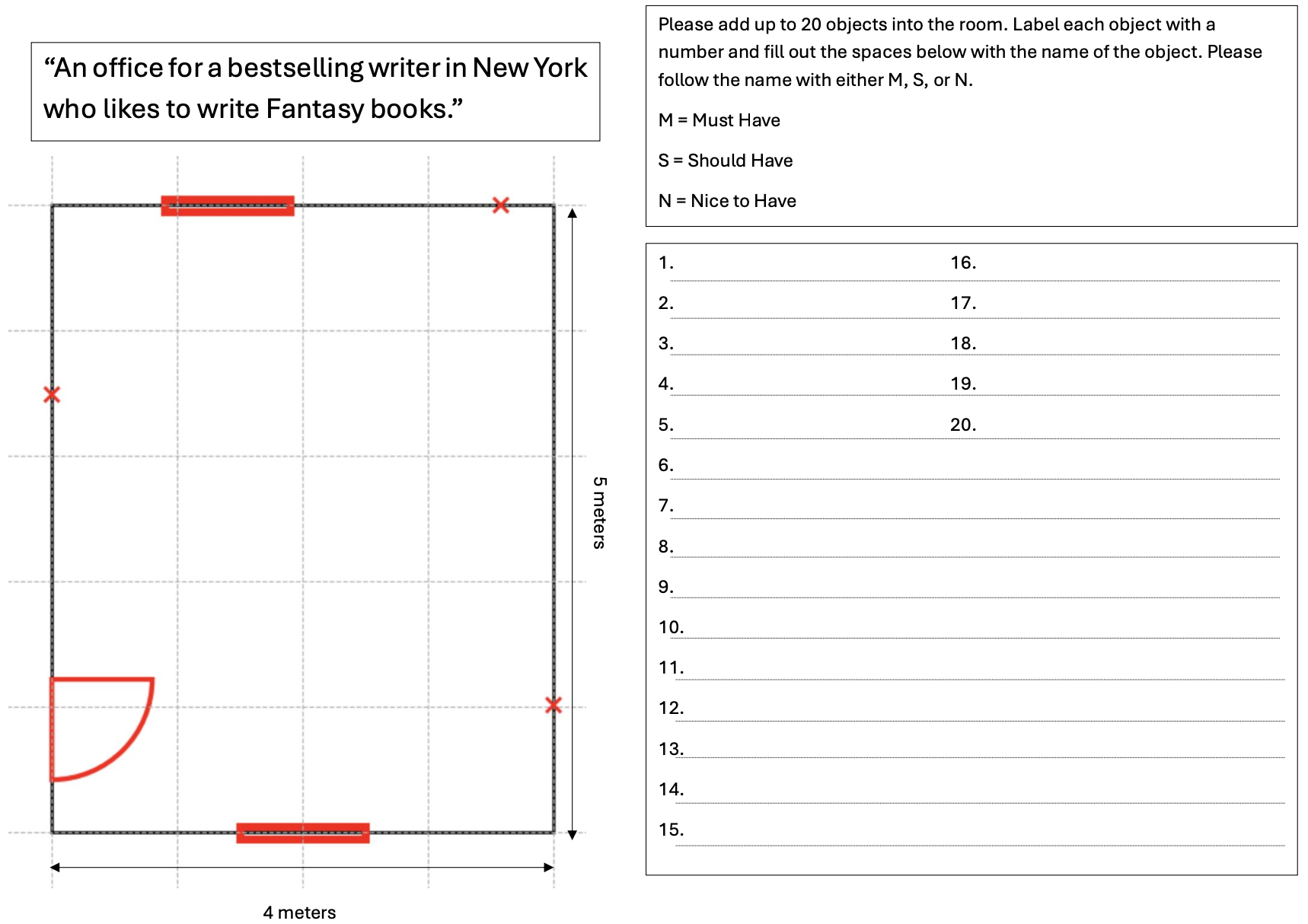}
\end{figure}

\begin{figure*}
\centering
\includegraphics[width = 0.3\textwidth]{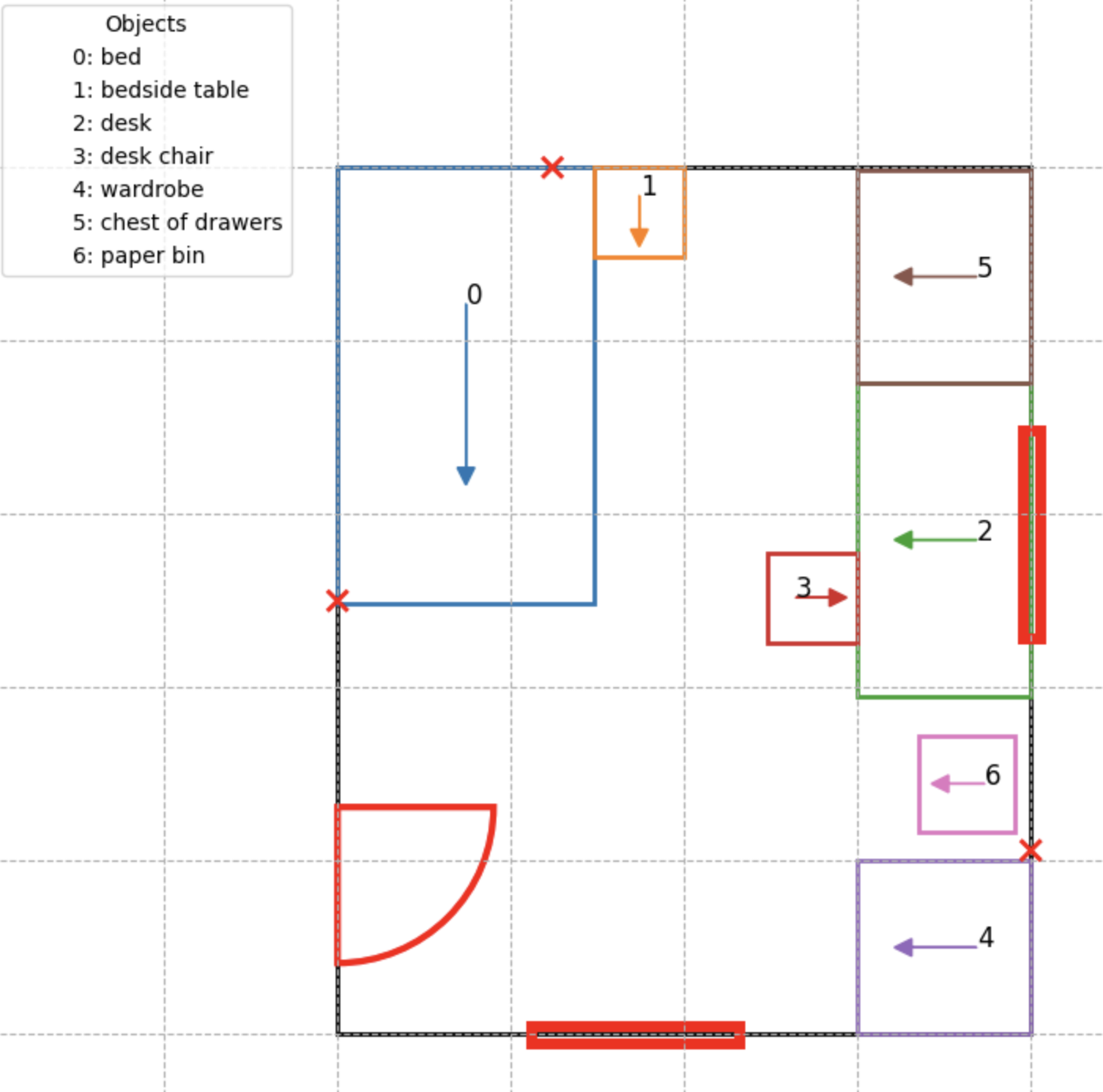}
\includegraphics[width = 0.3\textwidth]{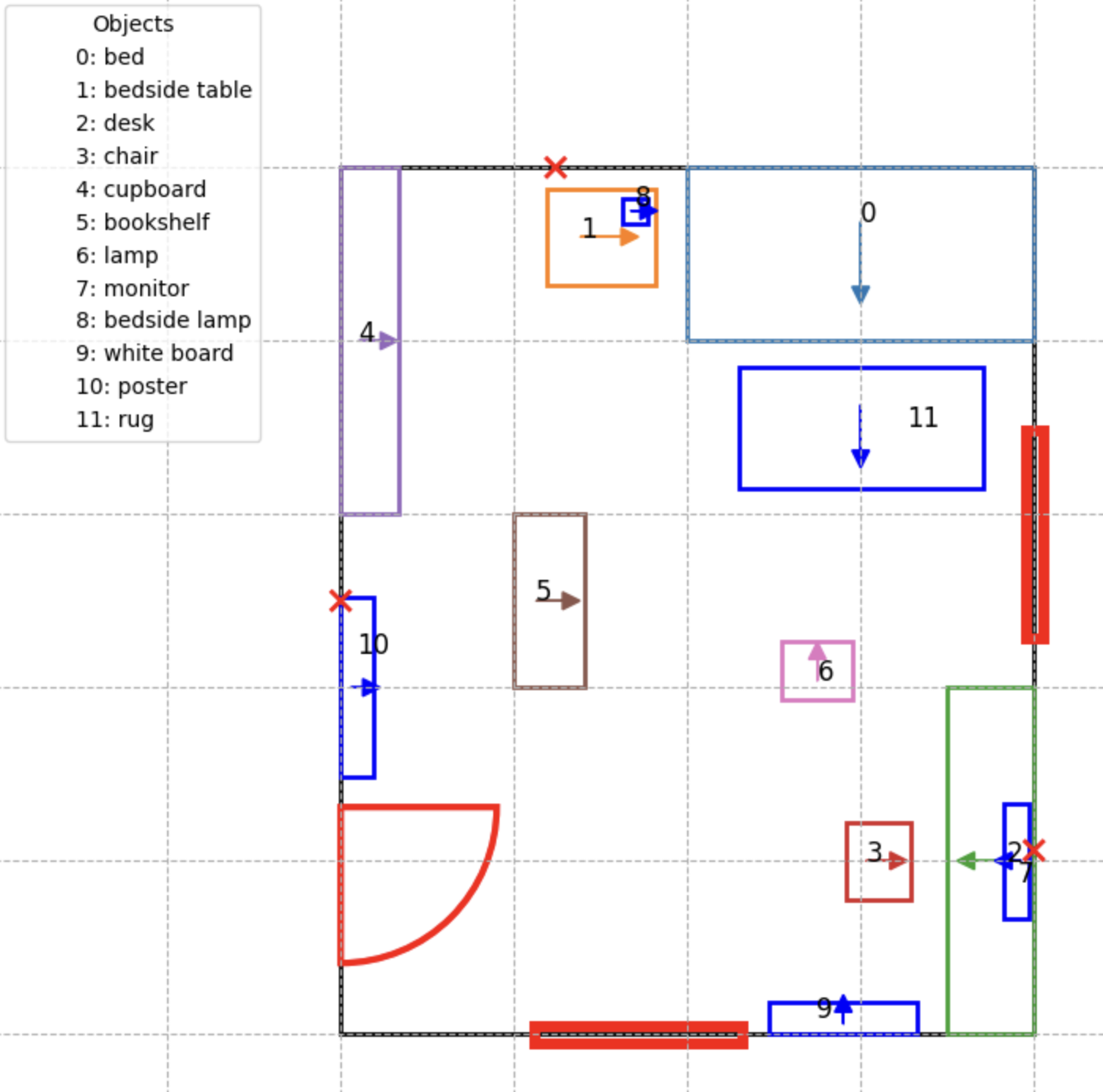}
\includegraphics[width = 0.3\textwidth]{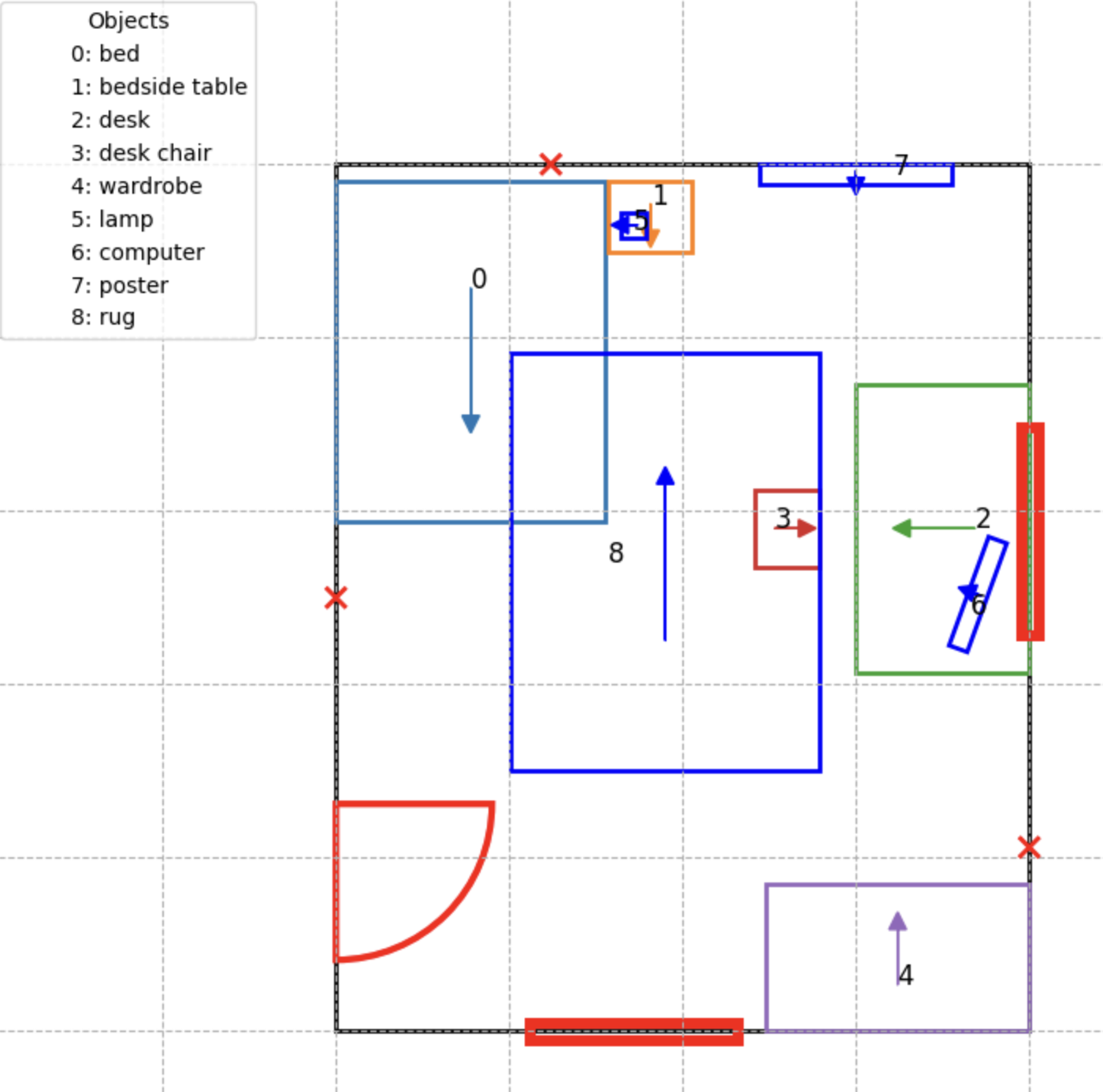}
\includegraphics[width = 0.3\textwidth]{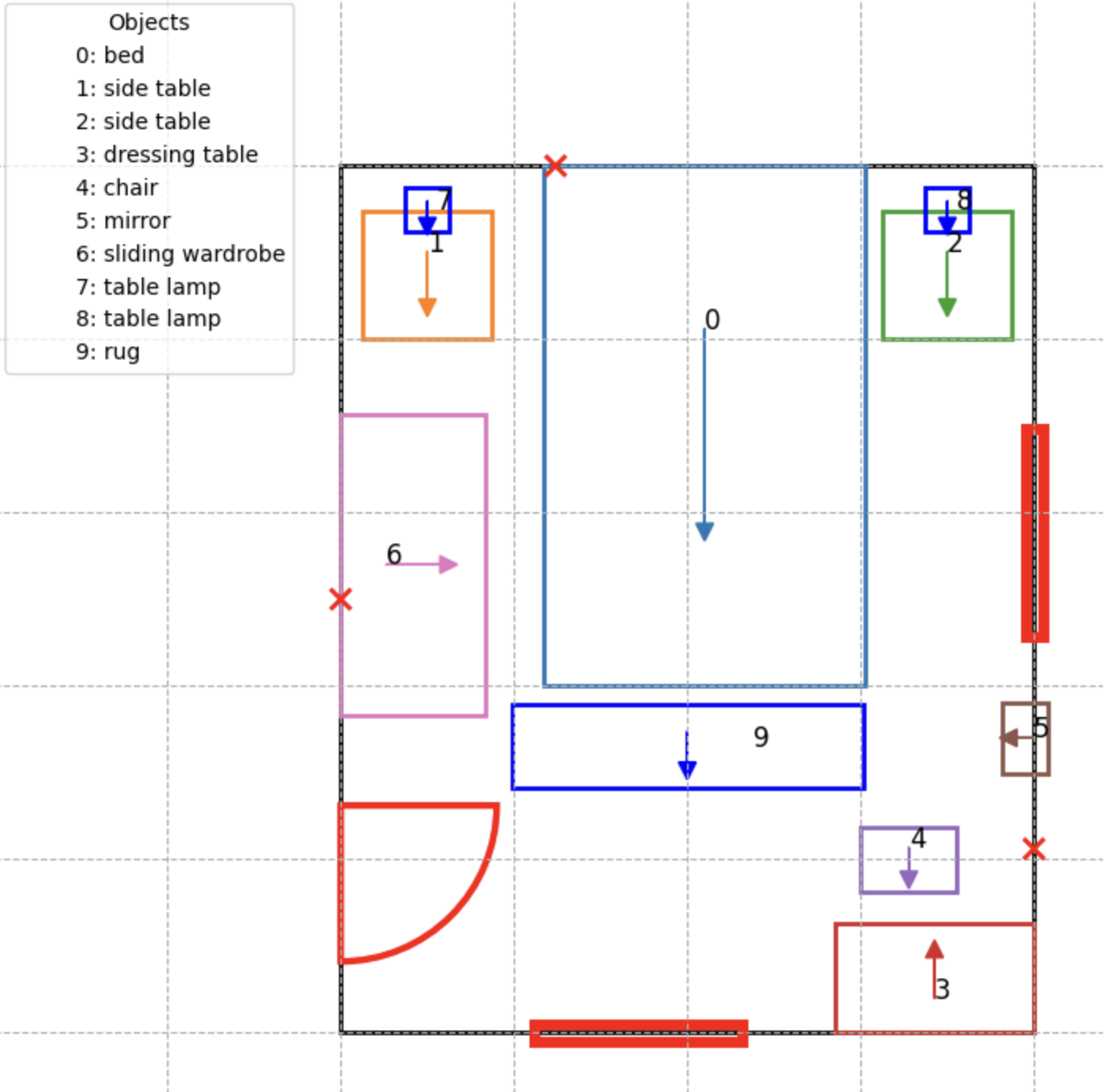}
\includegraphics[width = 0.3\textwidth]{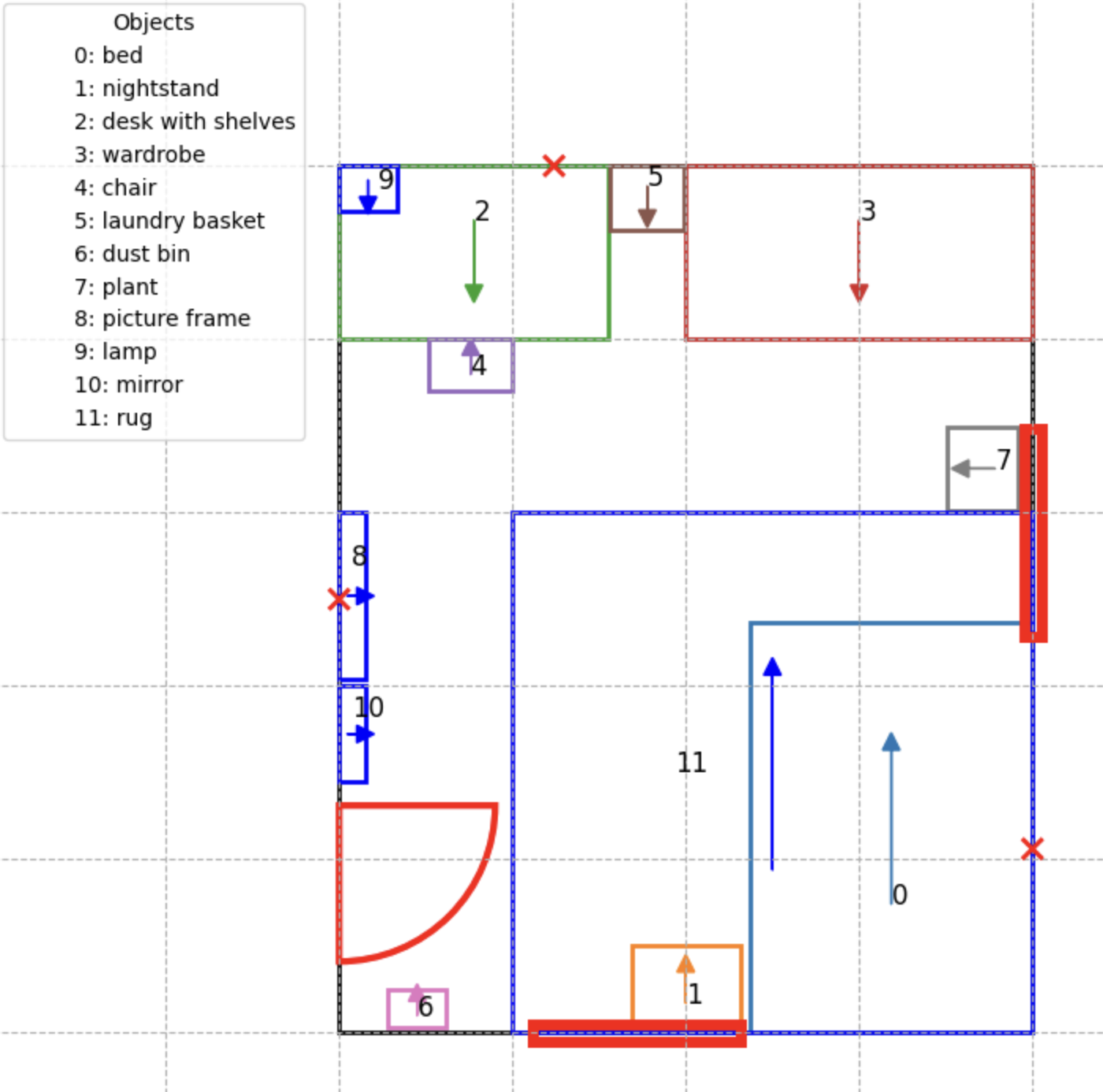}
\caption{Layouts designed by 5 novice human designers for the prompt: ``a bedroom that is 4m x 5m."}
\label{figure: h1}
\end{figure*}

\begin{figure*}
\centering
\includegraphics[width = 0.45\textwidth]{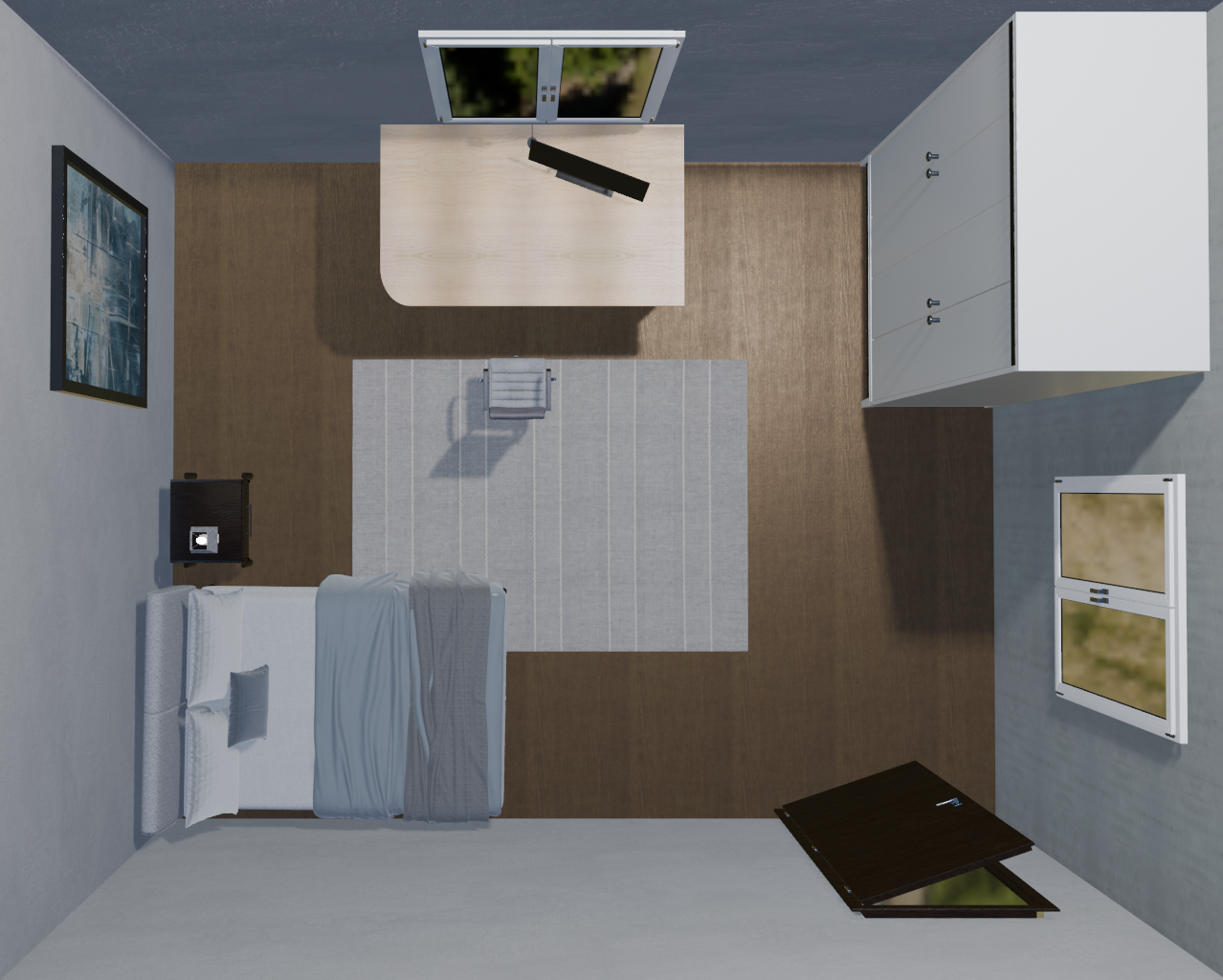}
\includegraphics[width = 0.45\textwidth]{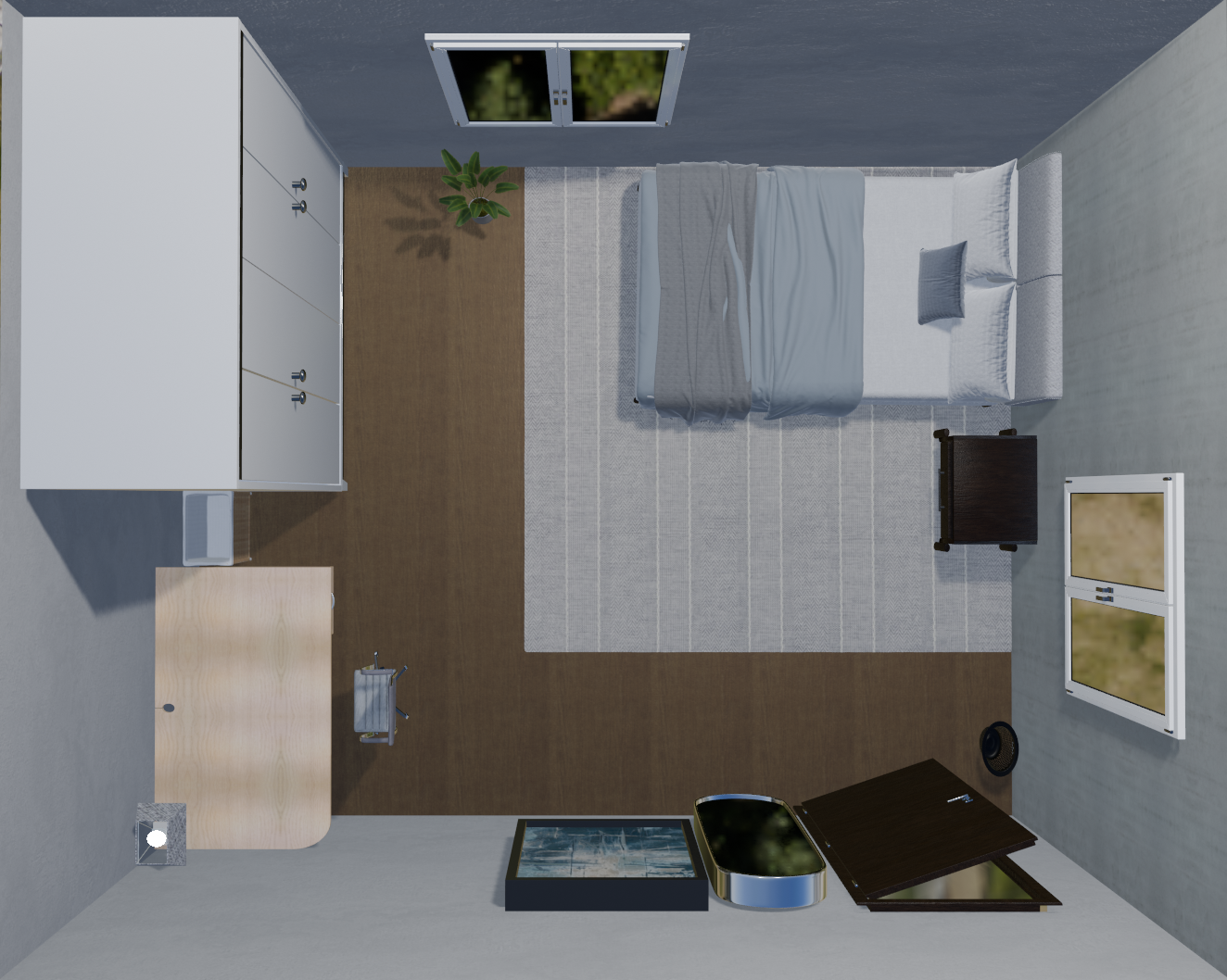}
\caption{Two layouts chosen from \autoref{figure: h1}, rendered in Blender \cite{Blender}, using assets from BlenderKit \cite{BlenderKit2024}.}
\end{figure*}

\begin{figure*}
\centering
\includegraphics[width = 0.3\textwidth]{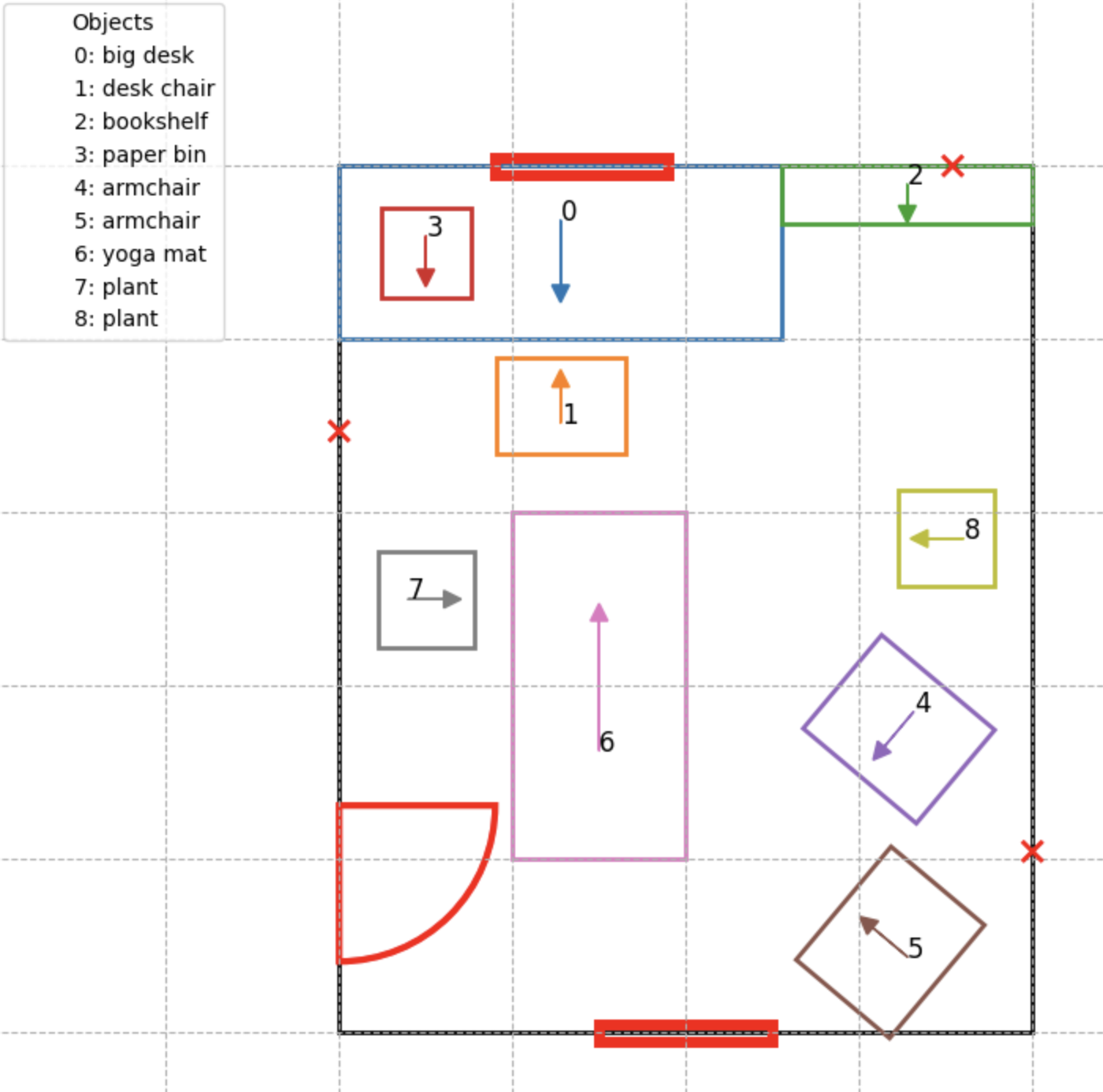}
\includegraphics[width = 0.3\textwidth]{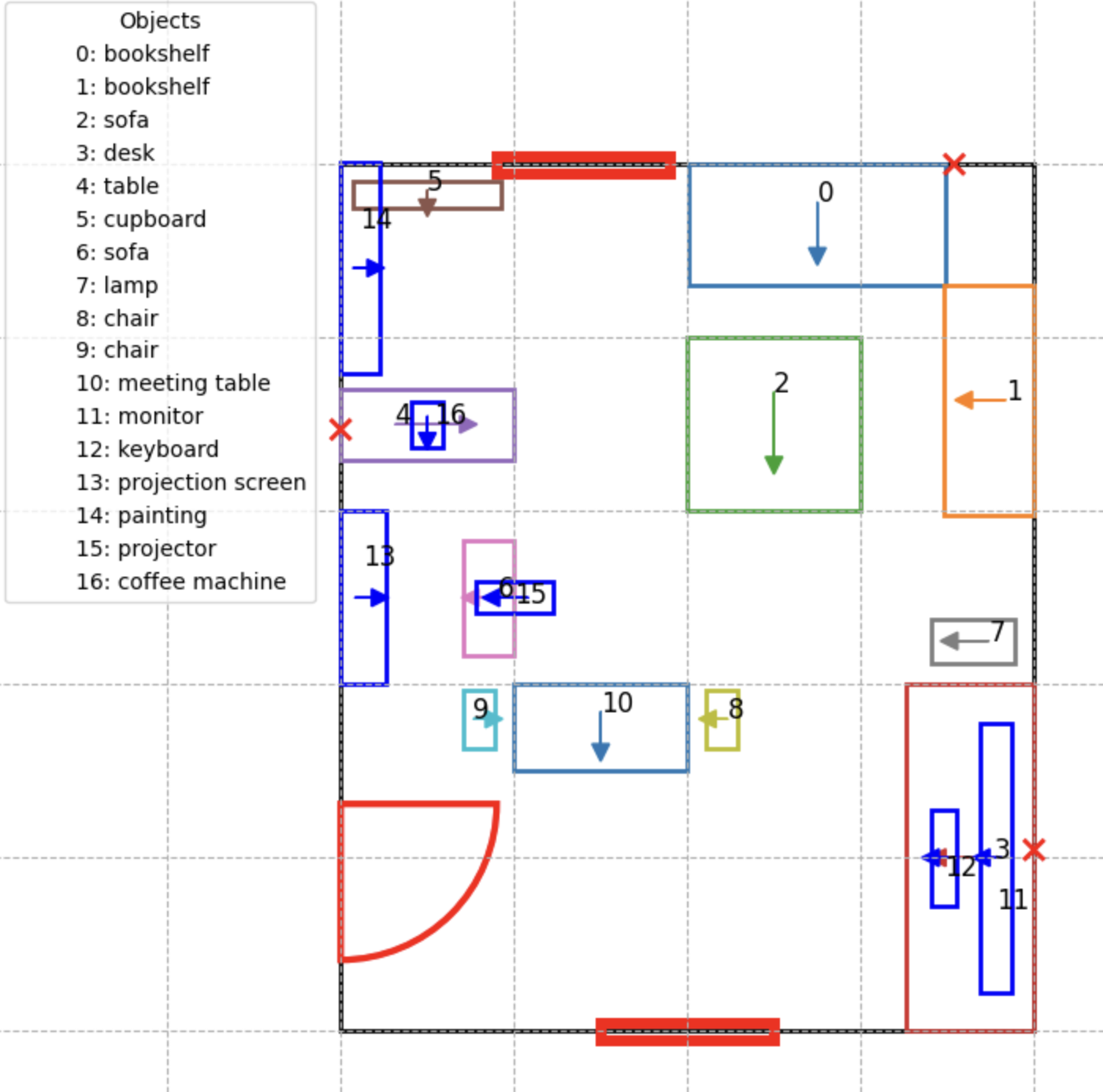}
\includegraphics[width = 0.3\textwidth]{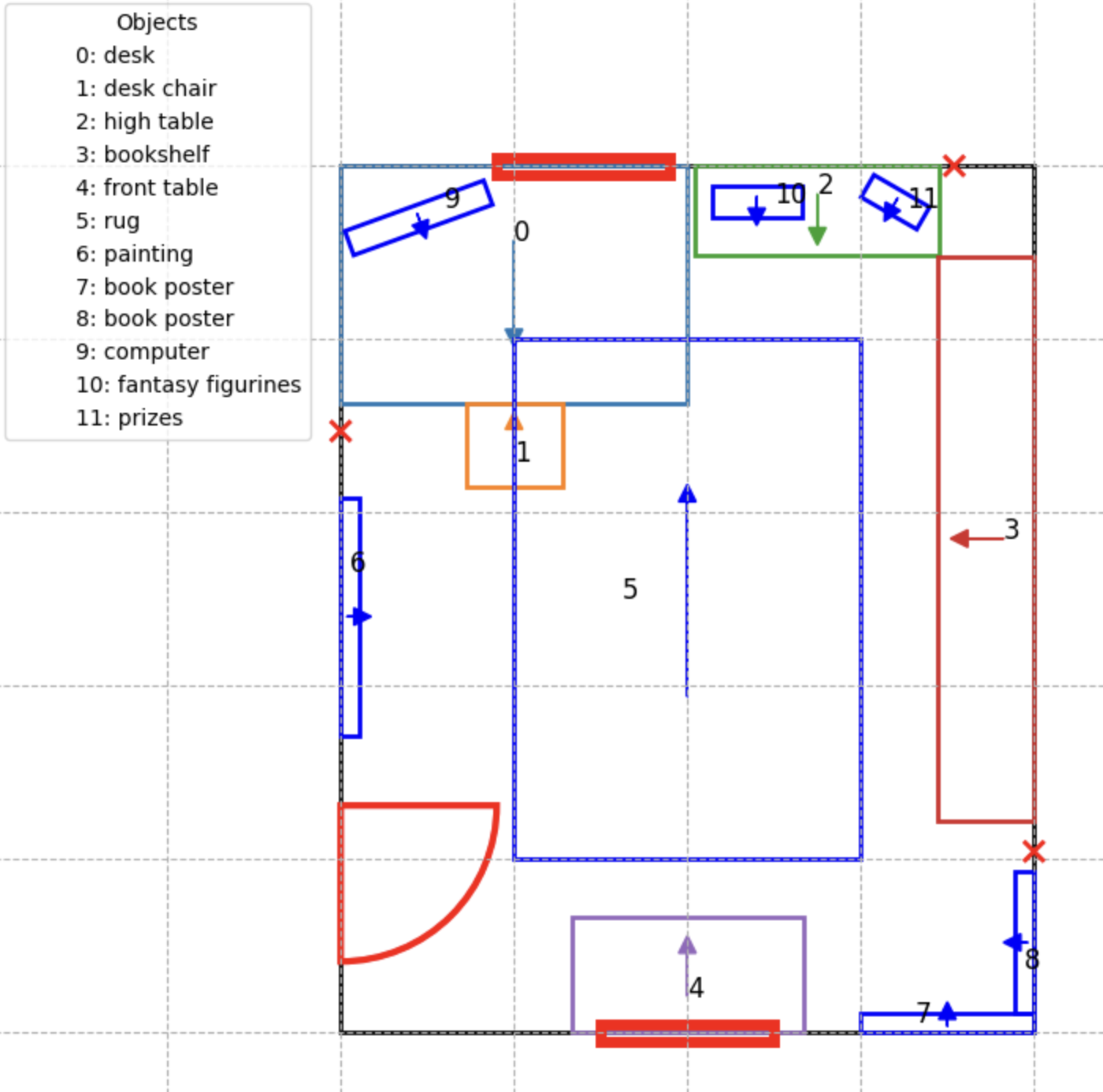}
\includegraphics[width = 0.3\textwidth]{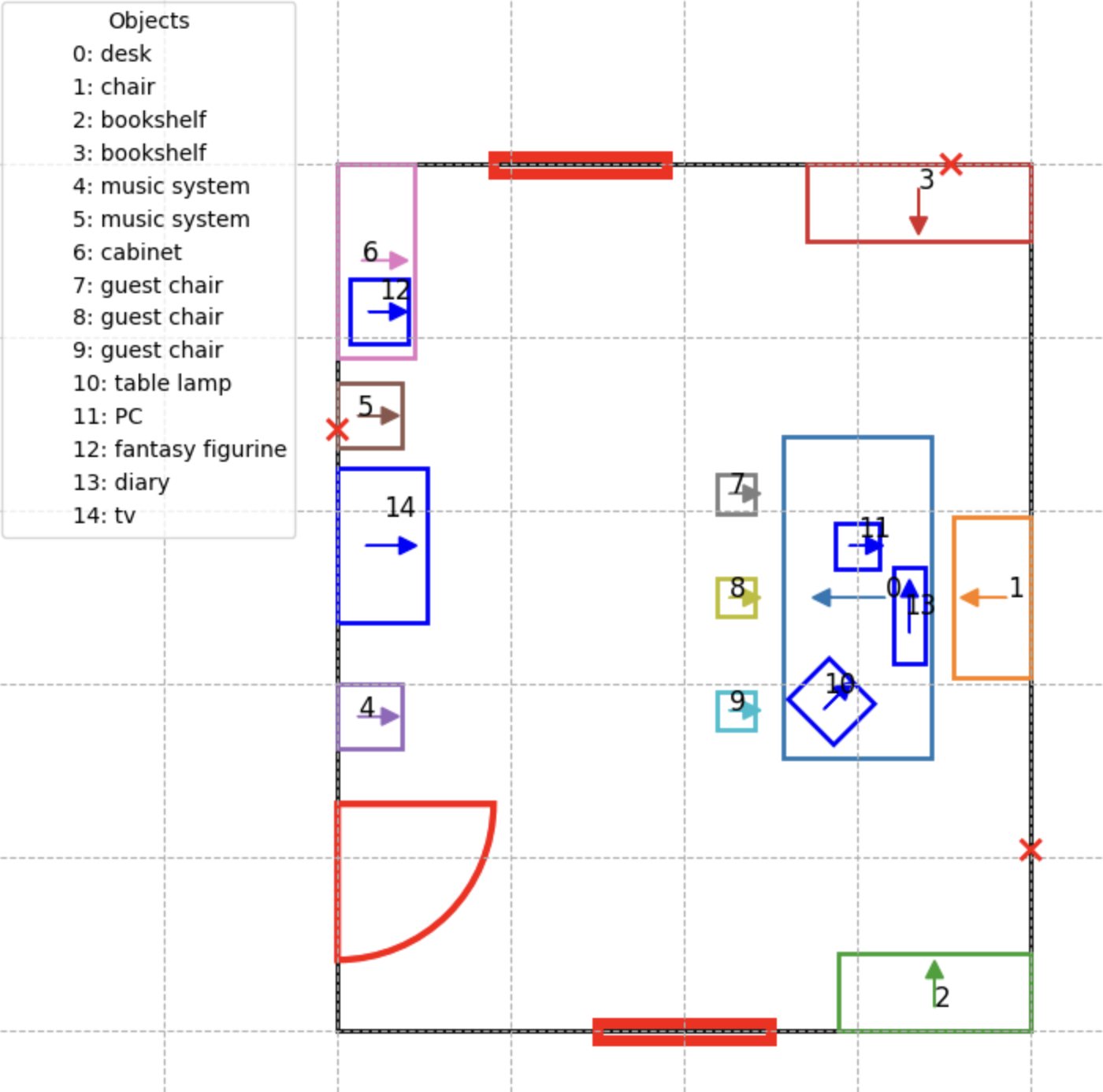}
\includegraphics[width = 0.3\textwidth]{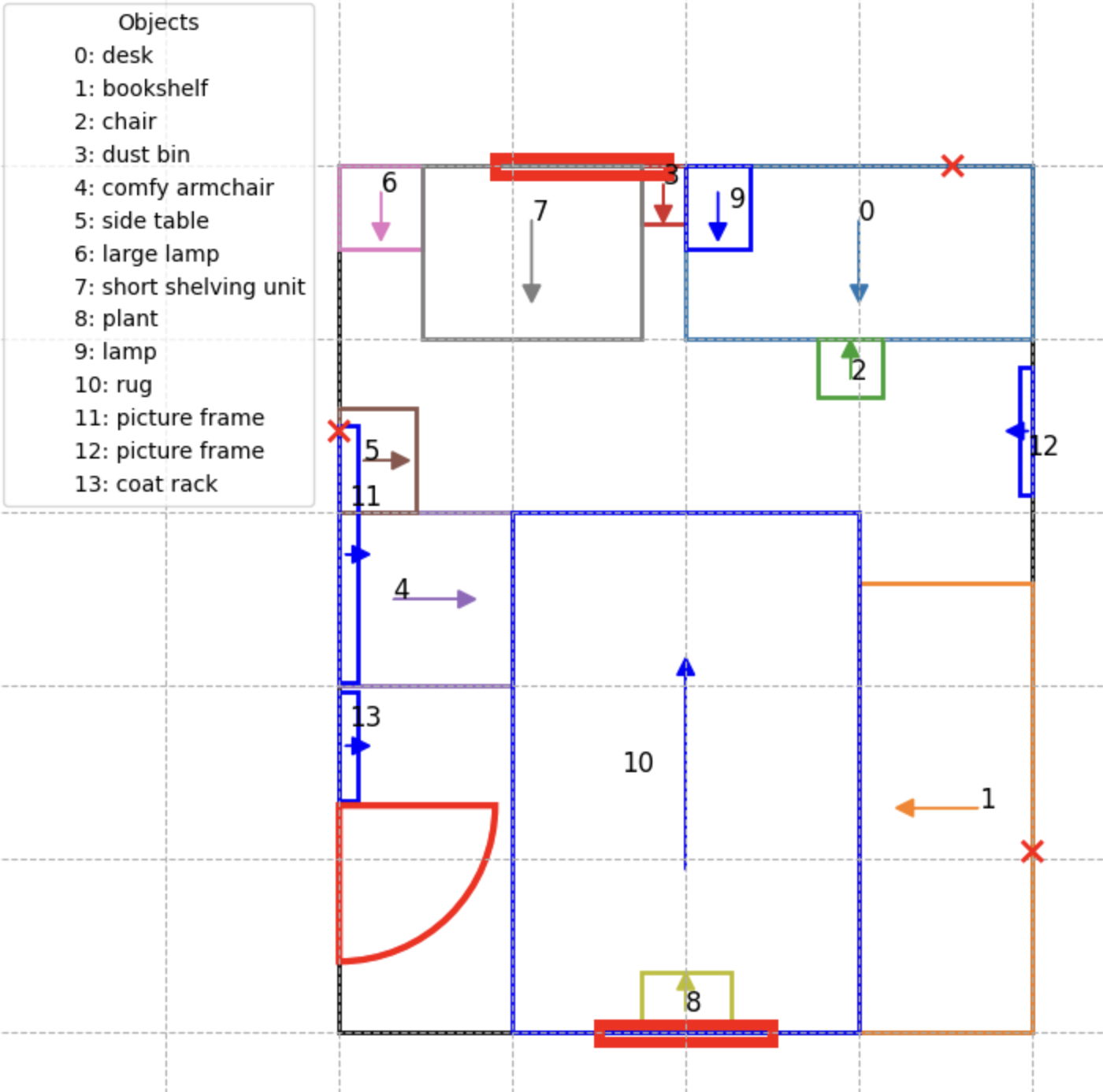}
\caption{Layouts designed by 5 novice human designers for the prompt: ``an office for a bestselling writer in New York who likes to write Fantasy books."}
\label{figure: h2}
\end{figure*}

\begin{figure*}
\centering
\includegraphics[width = 0.45\textwidth]{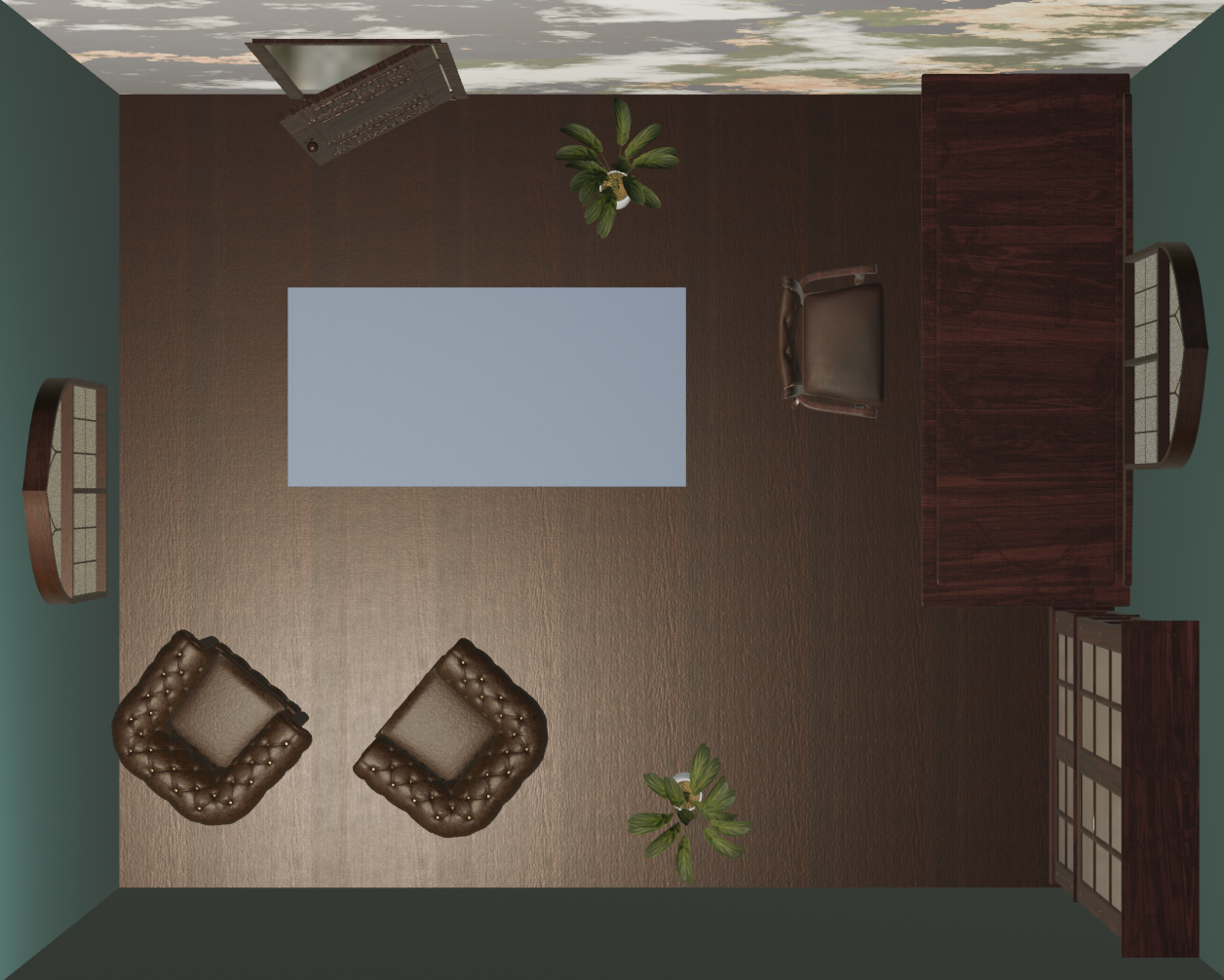}
\includegraphics[width = 0.45\textwidth]{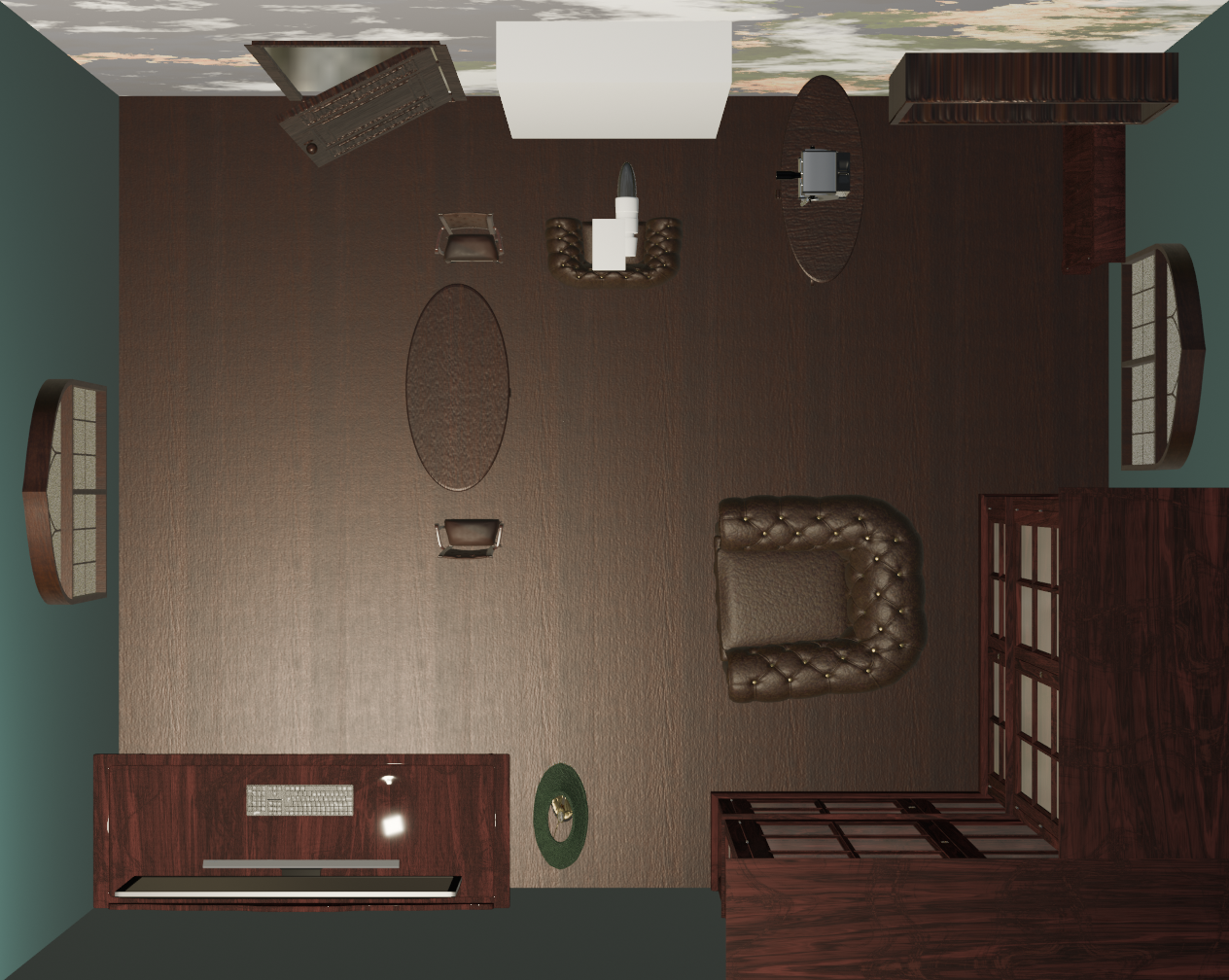}
\caption{Two layouts chosen from \autoref{figure: h2}, rendered in Blender \cite{Blender}, using assets from BlenderKit \cite{BlenderKit2024}.}
\end{figure*}

\newpage
\section{Blank Constraint Cost Functions}
\begin{minipage}{\textwidth}
    \includegraphics[page=1, scale = 0.7]{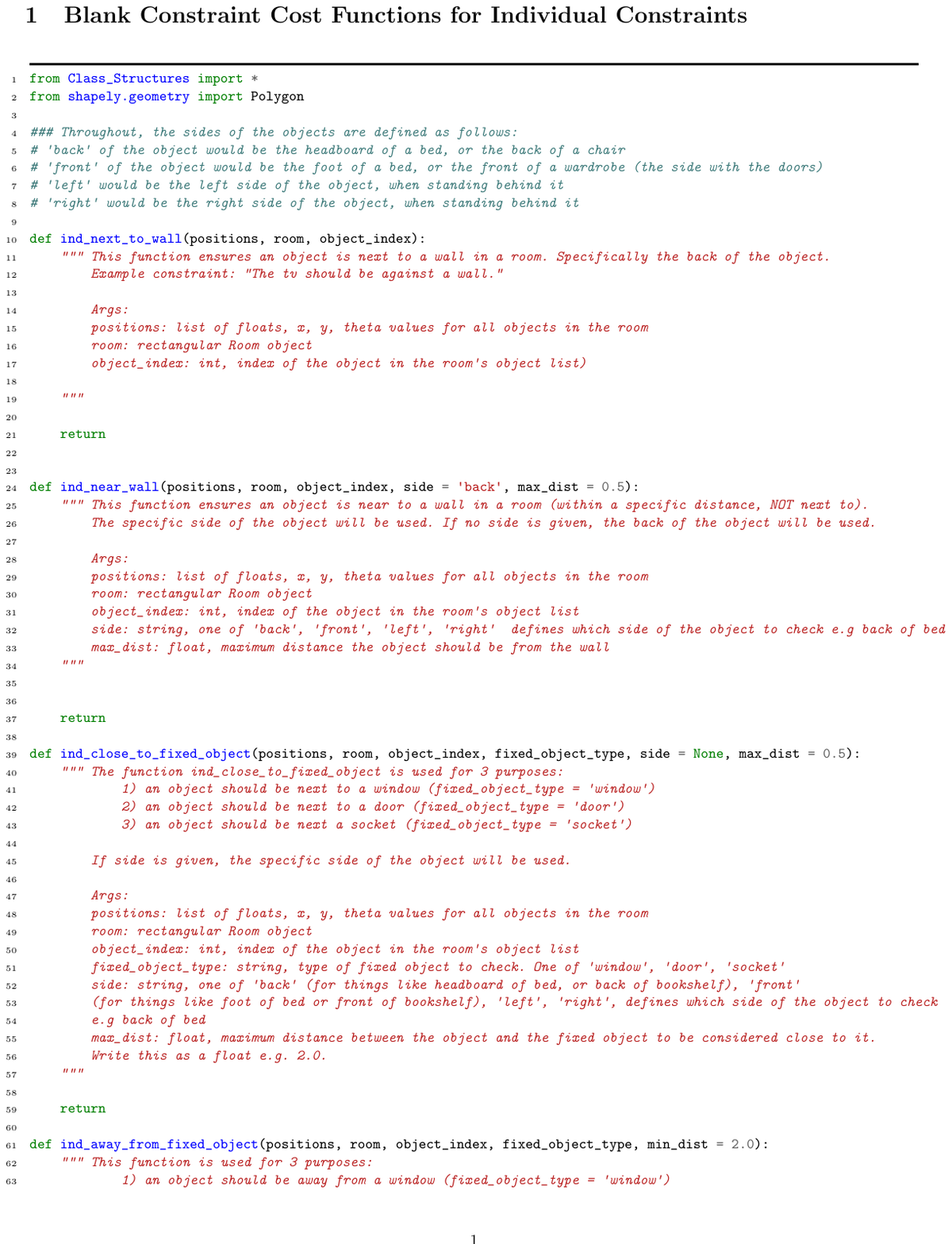}
\end{minipage}

\newpage
\includepdf[pages=2-, scale = 0.75]{supplementary/BlankConstraints.pdf}

\newpage
\section{Full example for ``A bedroom that is 4m x 5m.''}
\begin{minipage}{\textwidth}
    \includegraphics[page=1, scale = 0.7]{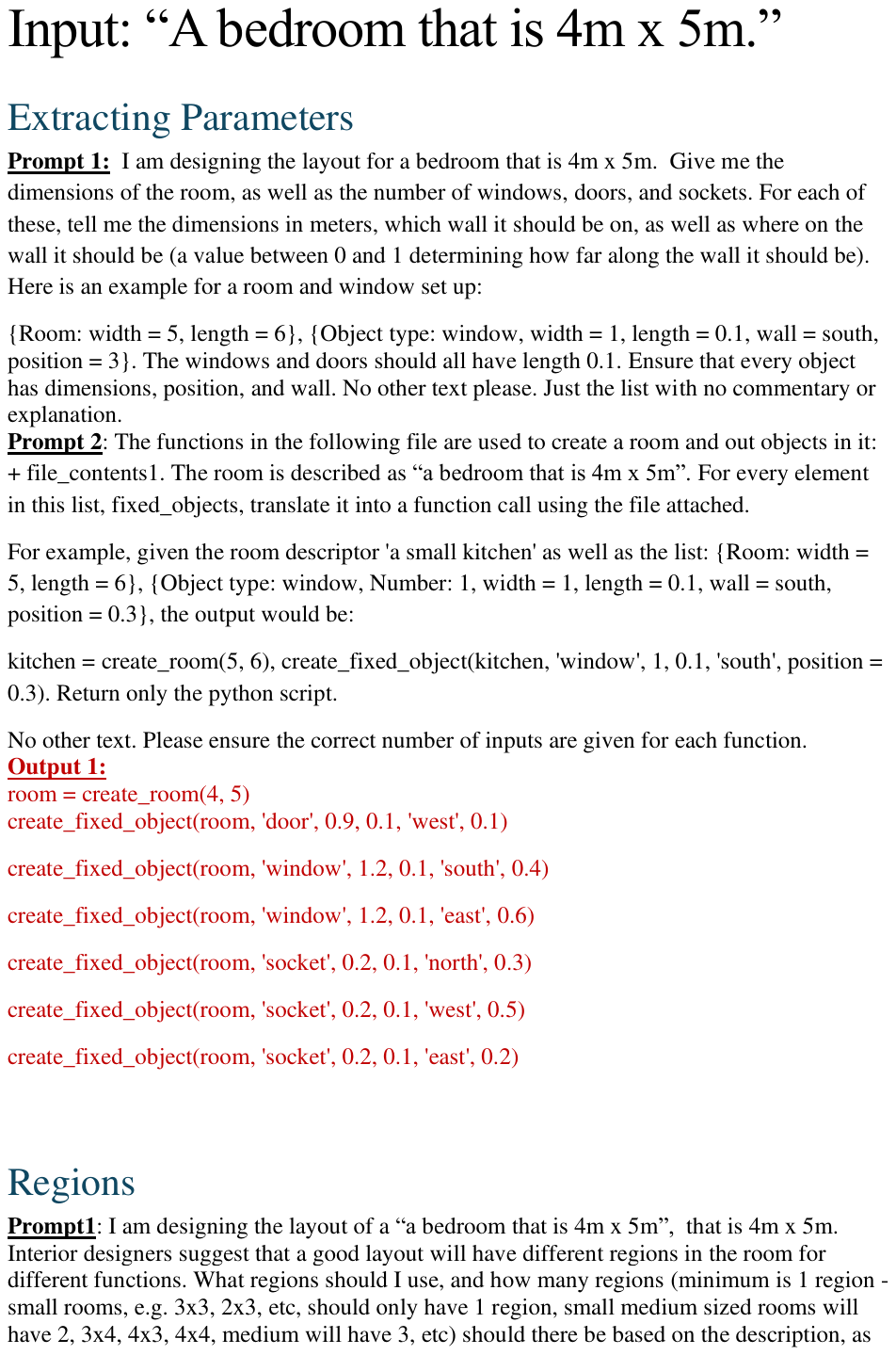}
\end{minipage}

\newpage
\includepdf[pages=2-, scale = 0.85]{supplementary/fulloutput.pdf}

\end{document}